\documentclass[useAMS,usenatbib]{mn2e}

\usepackage{graphicx}
\usepackage{color}

\title[Angular momentum as a second parameter]
  {Simulations of the formation and evolution of isolated dwarf
 galaxies $-$ II. Angular momentum as a second parameter}
\author[J. Schroyen et al.]
  {J.~Schroyen$^1$\thanks{JS thanks the Fund for Scientific Research -
  Flanders, Belgium (FWO) for financial support, E-mail:
  Joeri.Schroyen@UGent.be.}, S.~De Rijcke$^1$,
  S.~Valcke$^1$\thanks{Doctoral Fellow of the Fund for Scientific
  Research - Flanders, Belgium (FWO), E-mail:
  Sander.Valcke@UGent.be.}, A. Cloet-Osselaer$^1$, H.~Dejonghe$^1$\\ $^1$Sterrenkundig
  Observatorium, Ghent University, Krijgslaan 281, S9, 9000 Gent,
  Belgium}
\date{Accepted 2011 May 14. Received 2011 April 13; in original form 2010 December 14}

\pagerange{\pageref{firstpage}--\pageref{lastpage}} \pubyear{2010}

\def\LaTeX{L\kern-.36em\raise.3ex\hbox{a}\kern-.15em
    T\kern-.1667em\lower.7ex\hbox{E}\kern-.125emX}

\newcommand{\unit}[1]{\ensuremath{\, \mathrm{#1}}}

\begin{document}

\label{firstpage}

\maketitle

\begin{abstract}
We show results based on a large suite of N-Body/SPH simulations of
isolated, flat dwarf galaxies, both rotating and non-rotating. The main
goal is to investigate possible mechanisms to explain the observed
dichotomy in radial stellar metallicity profiles of dwarf galaxies:
dwarf irregulars (dIrr) and flat, rotating dwarf ellipticals (dE)
generally possess flat metallicity profiles, while rounder and
non-rotating dEs show strong negative metallicity gradients.

These simulations show that flattening by rotation is key to reproducing
the observed characteristics of flat dwarf galaxies, proving
particularly efficient in erasing metallicity gradients. We propose a
``centrifugal barrier mechanism'' as an alternative to the previously
suggested ``fountain mechanism'' for explaining the flat metallicity
profiles of dIrrs and flat, rotating dEs. While only flattening the
dark-matter halo has little influence, the addition of angular momentum
slows down the infall of gas, so that star formation (SF) and the
ensuing feedback are less centrally concentrated, occurring
galaxy-wide. Additionally, this leads to more continuous SFHs by
preventing large-scale oscillations in the SFR (``breathing''), and
creates low density holes in the ISM, in agreement with observations of
dIrrs.

Our general conclusion is that rotation has a significant influence on
the evolution and appearance of dwarf galaxies, and we suggest angular
momentum as a {\em second parameter} (after galaxy mass as the dominant
parameter) in dwarf galaxy evolution. Angular momentum differentiates
between SF modes, making our fast rotating models qualitatively resemble
dIrrs, which does not seem possible without rotation.

\end{abstract}


\section{Introduction}

Morphologically, dwarfs come in two broad classes. Early-type dwarfs, or
dwarf elliptical galaxies \citep[dE;]{ferbin}, are ``red and dead'' in
the sense that their stellar populations are predominantly old and that
they are usually not actively forming stars. They almost completely lack
the raw material for star formation: gas. In a small fraction of
dEs, low level central star formation continues at a rate of less than
one solar mass every $1000$ years \citep{rij03a, li06}. dEs with
luminosities below M$_V\sim -14$~mag are usually called dwarf
spheroidals, or dSphs. As a class, dEs are slowly rotating objects,
flattened by velocity anisotropy \citep{ge03}. Late-type dwarfs, or
dwarf irregular galaxies (dIrr; see e.g. \citealt{ski05}), are gas-rich
and are actively forming stars at a rate of about one solar mass every
$100-1000$ years. As a class, dIrrs are flattened by rotation
\citep{co00}. Noticeably, dSphs/dEs are found predominantly in
dense galactic environments while dIrrs are typically found in more
sparsely populated environments. This is the so-called
morphology-density relation \citep{bi87, co09}. In the Perseus cluster,
all dwarfs, irrespective of type, appear to avoid the very dense cluster
center \citep{pe09}. All this suggests that the environment is at least
to some extent responsible for many of the differences between dIrrs and
dEs.

Despite their quite different properties, the two types of dwarfs also
share many properties. They populate roughly the same mass, metallicity,
luminosity, flattening \citep{binggeli:flatdist} and size regimes and
they have, to a good approximation, exponentially declining
surface-brightness profiles. Moreover, the ``boundaries'' between the
dwarf classes are not clear-cut and transition-type objects with mixed
properties exist \citep{gre03}. This body of data provides us with
evidence for evolutionary links between, or at least a ``common
ancestry'' for, the different types of dwarfs. As shown by
e.g. \citet{ma06}, the combined action of ram-pressure stripping and
tidal stirring on a star-forming, rotating late-type dwarf entering the
halo of a Milky Way-like massive galaxy can remove most of its gas and
angular momentum, effectively transforming it into a quiescent,
non-rotating early-type dwarf.

Dwarf galaxies entering a dense environment that are affected by
ram-pressure stripping but not (or much less so) by tidal stirring,
would be expected to keep many of their late-type structural
properties and one would expect to find dIrr/dE transition-type
dwarfs. Indeed, quiescent dwarfs have been observed that are
significantly more flattened and faster rotating than the average dE
\citep{rij03b, vanzee04}, contain gas and dust \citep{co03, buy05,
  lo10}, and often host embedded stellar disks and spiral structures
\citep{je00, ba02, rij03b, gra03}.

\subsection{Metallicity profiles}

In general, dIrrs display chemical homogeneity practically throughout
their entire stellar and gaseous bodies
\citep{tolstoy:dgreview,kobulnicky-skillman:dirrmetprof}. The Small
Magellanic Cloud \citep[SMC;][]{dufour-harlow:smc,pagel:smc}, NGC 6822
\citep{hernandez:ngc6822}, and Sextans~A \citep{kaufer:sextansA} are
examples of dIrrs without a significant chemical or abundance gradient
in their gas content; the SMC \citep{cioni:smc} and IC 1613
\citep{bernard:ic1613} also lack a stellar metallicity gradient. Thus, a
flat radial metallicity profile seems to be a rather general
characteristic of dIrrs. \citet{koleva-sven:demetprof} present radial
stellar metallicity profiles, derived from optical VLT spectra, of a
sample of 16 dEs belonging to the Fornax cluster and to nearby groups of
galaxies. They find that ten of those, predominantly round and
non-rotating, show a strong negative metallicity gradient. The six most
flattened and most strongly rotating galaxies in the sample, however,
show no significant gradient:~like the rotationally flattened dIrrs,
they are chemically homogeneous. Previous studies have also
predominantly found negative metallicity gradients for dEs in the Local
Group (e.g. \citealt{harbeck01}; \citealt{alard01}, Saggitarius; - the
DART project: \citealt{tolstoy04}, Sculptor; \citealt{battaglia06},
Fornax; \citealt{battaglia10}, Sextans), around M81 \citep{lianou10} and
in the Coma cluster \citep{brok2011}.

The findings of \citet{koleva-sven:demetprof} suggest that, while
total mass is most likely the dominant factor \citep[as is concluded
from the simulations of:][]{sander:dgmodels, sa10, re09, sti07}, angular
momentum is an important second parameter in the chemical evolution of
dwarf galaxies:~fast rotating dwarf galaxies show a tendency to be
chemically much more homogeneous than dwarfs with slow or no rotation.

An often quoted means of erasing metallicity gradients in flattened
dwarf galaxies is the so-called ``fountain mechanism'' \citep[for
example :][and references
therein]{maclow:bo,tolstoy:bo,deyoung90,deyoung:fountain,binggeli}. The
idea behind this mechanism is that the supernova feedback of a
centralized star-formation event is capable of ejecting significant
amounts of hot, enriched gas through a cavity or ``chimney'' along the
galaxy's minor axis. Subsequently, part of this gas can rain back down
on the galaxy's disk, as in a fountain, diluting any metallicity
gradient that might be present. In round low-mass galaxies, centrally
concentrated supernova feedback is expected to ``blow away'' {\em all}
the gas rather than to ``blow out'' only the enriched hot gas. If this
fountain mechanism is correct, the absence or presence of a metallicity
gradient is determined by two parameters: a dwarf galaxy's mass and its
flattening or geometry. 

Alternatively, due to the ``centrifugal barrier'' in a rotating
galaxy, gas cannot readily flow to the center and build up a strong
centrally concentrated star formation event. One would therefore
expect that rotation will naturally lead to more spatially extended
star formation and thus to more spatially homogeneous stellar
populations. In a similar vein, angular momentum has been proposed in
the literature as the fundamental parameter setting low angular
momentum starbursting Blue Compact Dwarfs apart from the more
continuously star-forming high angular momentum dIrrs
\citep{vanzee01}.

\subsection{Paper}

In this paper, we use a suite of new N-body/SPH simulations to
investigate how flattening affects the star-formation histories and
chemical evolution of the isolated dwarf galaxy models presented in
\citet{sander:dgmodels}. Other N-body/SPH simulations of similar
star-forming, gas-rich dwarf galaxy models, though not always isolated,
have been performed by
e.g. \citet{pelu04,sti06,sti07,re09,governato10,sa10,sa11}. We flatten
our originally spherically symmetric models in different ways by
adapting their initial conditions, with and without adding rotation, and
compare the results both with the spherically symmetric originals and
with the available observations. Our main goal is to contrast the
``fountain mechanism'' with the ``centrifugal barrier'' hypothesis, and
to see if it's possible to produce dwarf galaxies with flat metallicity
profiles in isolation.

In section \ref{section:code}, we give a brief description of the
simulation code, followed by a description of the simulations themselves
in section \ref{section:simulations}. We present an analysis of the
simulations in section \ref{section:analysis}, discuss the results in
section \ref{section:results} and conclude in section
\ref{section:conclusion}.

\section{Codes}
\label{section:code}

For this research, we relied mainly on two codes: the Nbody-SPH code
{\sc Gadget-2} \citep{springel05} for the simulations, and our own home-made
analysis tool {\sc Hyplot} for analysing and visualising these
simulations. These will both briefly be described below.

\subsection{Astrophysical mechanisms}

The code we actually use for our simulations is a modified version of
the Nbody-SPH code {\sc Gadget-2} \citep{springel05}. The freely available
version only incorporates gravity and hydrodynamics, so to prove useful
for investigating dwarf galaxy formation and evolution a number of
additions were made. These include star formation, feedback and
radiative cooling. Re-ionization or a UV background is not included
in our models. Below we give a brief overview of the implementations,
more detailed information can be found in
\citet{sander:dgmodels,sander:sph} and in \citet{sander:phd}.

\subsubsection{Star formation}
\label{section_SF}

Stars are formed when three criteria are satisfied:
\begin{eqnarray}
 \rho_{g} & \geq & \rho_{c} = 0.1~ \mathrm{cm^{-3}} \\
 T & \leq & T_{c} = 15000 \mathrm{K} \\
 \vec{\nabla}.\vec{v} & \leq & 0 .
\end{eqnarray}
So we have a density treshold, a temperature treshold and the
requirement that the gas be converging. We do not explicitly implement a
Jeans criterion. Gas particles eligible for star formation are turned
into stars according to the Schmidt law \citep{schmidt59}:
\begin{equation}
\frac{\mathrm{d}\rho_{s}}{\mathrm{d}t} =
 -\frac{\mathrm{d}\rho_{g}}{\mathrm{d}t} = c_{\star}\frac{\rho_{g}}{t_{g}},
\end{equation}
where $\rho_{s}$, $\rho_{g}$ and $c_{\star}$ are, respectively, the
density of stars and gas, and a dimensionless star formation efficiency
factor. $t_{g}$ is taken to be the dynamical time for the gas $1/\sqrt{4
\pi G \rho_{g}}$.

\subsubsection{Feedback}

Produced star particles are represented as ``simple stellar
populations'' (SSP), applying the Salpeter IMF for the probability that
a star of mass $m$ resides in the SSP:
\begin{equation}
\Phi(m)\mathrm{d}m = Am^{-(1+x)}\mathrm{d}m,
\end{equation}
with $x=1.35$, $A=0.06$, and the limits for stellar masses are
$m_{\mathrm{l}} = 0.01~\mathrm{M_{\odot}}$ and $ m_{\mathrm{u}} =
60~\mathrm{M_{\odot}}$.  Feedback from a star particle is given through
stellar winds (SW) and supernovae (SN, type II and Ia), and includes the
return of both energy and enriched gas to the ISM (thermal
feedback). These are transferred to the surrounding gas particles
according to the SPH smoothing kernel. The lower limits for SNIa and
SNII respectively are $3 \mathrm{M_{\odot}}$ and $8
\mathrm{M_{\odot}}$, upper limits $8 ~ \mathrm{M_{\odot}}$ and $60
\mathrm{M_{\odot}}$, and progenitor lifetimes are $5.4 \times 10^{6}$ yr
and $1.5 \times 10^{9}$ yr.

The energy feedback for both types of SN is taken to be $10^{51}$ erg,
for SW this is $10^{50}$ erg, which are all transferred to the ISM with
an efficiency of $0.1$.

The returned mass fractions for SNIa and SNII are $0.00502$ and $0.112$,
and the metal yields from these supernovae are taken from, respectively,
\citet{travaglio04} (their b20\_3d\_768 model) and
\citet{tsujimoto95}. From the last authors we also adopt
$N_{SNIa}/N_{SNII} = 0.15$ to set the fraction of stars in the relevant
mass range that reside in binary systems and that can go SNIa. The yield
$M_{i}$ of element $i$ by e.g. SNII is then calculated as:
\begin{equation}
M_{i} = M_{SSP} \frac{\int_{m_{\mathrm{SNII,l}}}^{m_{\mathrm{SNII,u}}} M_{i}(m) \Phi(m) \mathrm{d}m }{\int_{m_{\mathrm{l}}}^{m_{\mathrm{u}}} m \Phi(m) \mathrm{d}m }.
\end{equation}
The formula for SNIa is similar but simpler, since there is no
dependence on progenitor mass ($M_{i}(m)$ becomes $M_{i}$).

\subsubsection{Cooling}

Metal dependant gas cooling is implemented according to the cooling
curves from \citet{suthdop}. These curves are interpolated in
metallicity and temperature to obtain the cooling strength for a gas
particle. They are allowed to cool in this fashion to a minimum of
$10^4$K, below which further cooling is only possible through adiabatic
expansion.  

\subsection{Analysis and visualisation}

For the analysis we used our own {\sc Hyplot} package, which is freely
available on
SourceForge\footnote{http://sourceforge.net/projects/hyplot/}. It is an
analysis/visualisation tool specially suited for Nbody-SPH simulations
(currently only specifically for {\sc Gadget-2} datafiles), written mainly in
Python and C++. PyQT and Matplotlib are used for the GUI and the
plotting, and it is also fully scriptable in Python. All analysis, plots
and visualisations in this paper have been made using {\sc Hyplot}.

\section{Simulations}
\label{section:simulations}

In this section we will describe the simulations themselves: the models
used for the basic initial conditions, the additional setup we need for
our goals, and finally the grid of our production runs, together with a
preliminary evaluation of those runs to have an idea of the simulated
objects we have at our disposal.

\subsection{Initial conditions}

We base the initial conditions of our flattened dwarf galaxy
simulations on the spherically symmetric dwarf galaxy models of
\citet{sander:dgmodels}. We describe these briefly below and in Table
\ref{DG_models}. We introduce flattening into the models by adding
initial flattening and/or rotation. The precise way in which this is
done is described below. In the end, we have a set of flattened dwarf
galaxies, both rotating and non-rotating. This way, we can distinguish
between the effects of the \textit{geometry} (flattening only) and the
\textit{kinematics} (i.e. rotation).

\subsubsection{Spherically symmetric dwarf galaxy models}
\label{section:basic_models}

The basic spherically symmetric dwarf galaxy models come from
\citet{sander:dgmodels}. They consist of
\begin{enumerate}
 \item a dark matter halo with a cored \textit{Kuz'min Kutuzov}
   density profile \citep{kuzmin},
 \item a spherically symmetric homogeneous gas cloud, set to a
       fixed initial temperature of $10^{4} \mathrm{K}$, initial
       metallicity of $10^{-4} ~\mathrm{Z_{\odot}}$ and the gas
       particles initially at rest.
\end{enumerate}
We start off with only gas and dark matter (DM). The gas cools and
collapses into the DM gravitational potential well. Our modified version
of {\sc Gadget-2}, as described in the previous section, then allows the
gas particles to produce star particles if the criteria are satisfied
(of which the density criterion proves to be the most
important). Feedback of energy and of newly synthesized elements from
these star particles is accounted for. For specific information and
details on these models, we refer to Table \ref{DG_models} and
\citet{sander:dgmodels}. The actual initial condition files are
created by random (Poisson) sampling of these density profiles for both
components.

\begin{table}

 \caption{Details of the basic spherical dwarf galaxy models (see
 \citealt{sander:dgmodels}). Initial masses for the DM halo and gas are
 in units of $10^{6} \mathrm{M_{\odot}}$, radius in kpc.}

 \label{DG_models}

\begin{tabular}{l@{\extracolsep{10ex}}ccc}
\hline                                
$\mathrm{model}$ & $M_{DM,i}$ & $M_{g,i}$ & $a_{0}$\\ 
\hline
C01 & 206 & 44 & 0.439 \\ 
C03 & 330 & 70 & 0.513 \\ 
C05 & 660 & 140 & 0.646 \\ 
C07 & 1238 & 262 & 0.797 \\ 
C09 & 2476 & 524 & 1.004 \\ 
 
\hline 
\end{tabular}

\end{table}

\subsubsection{Adding initial rotation}

Rotation is added \textit{to the gas only} in the initial
conditions. The DM halo, implemented as a live halo, simply
provides a background gravitational potential and is not given any
rotation. Every gas particle is given a tangential velocity according to
the desired rotation profile. We align the $z$-axis with the galaxy's
rotation axis.

\begin{figure}
 \includegraphics[width=0.45\textwidth]{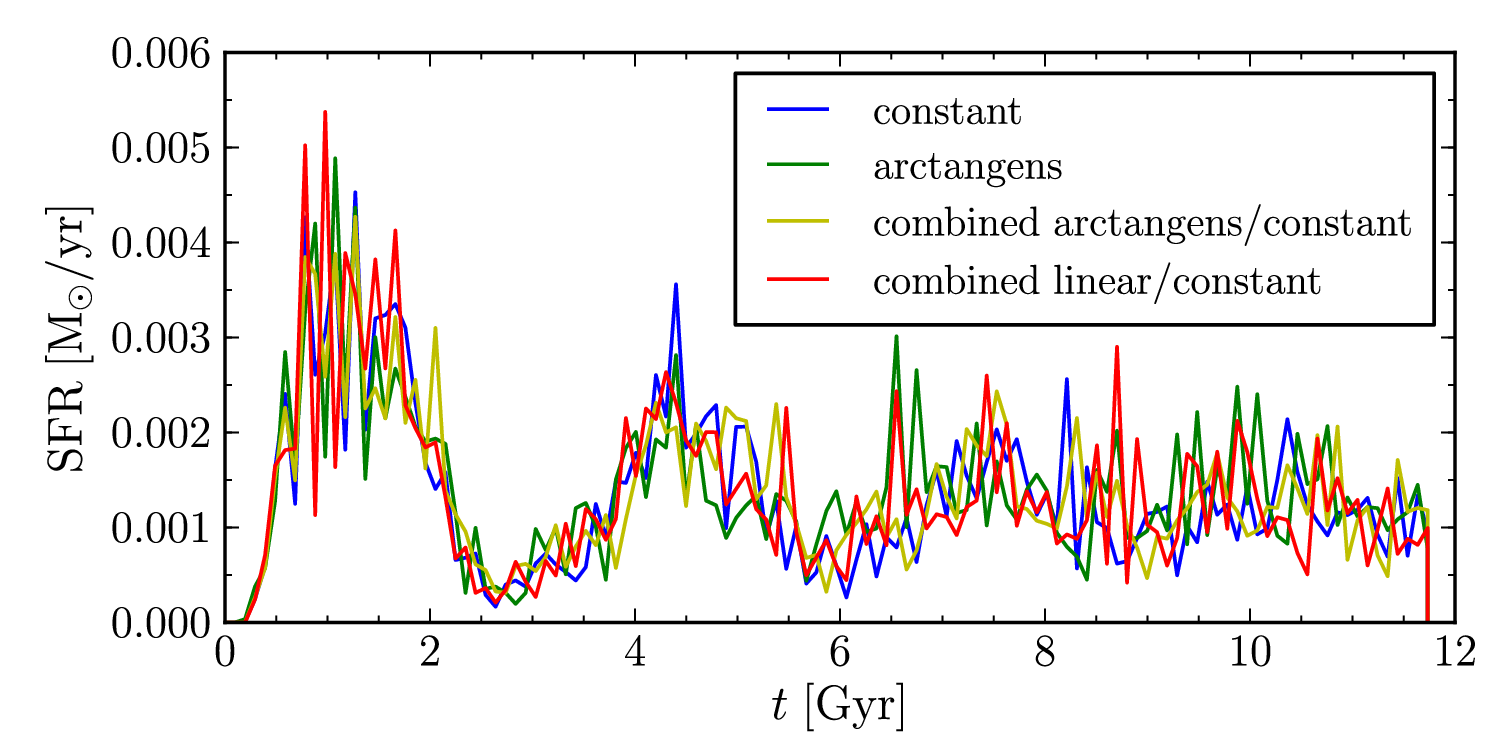}
 \caption{The star formation rate as a function of time for four
 different rotation profiles (see text for details).}  \label{rotprof_check}
\end{figure}

Our preferred rotation profile is a \textit{constant radial rotation
profile} (CR), which means that the net velocity given to each gas
particle is independent of radius. We do not use a \textit{solid body
rotation profile} (SBR), in which the velocity depends linearly on the
radial distance from the rotation axis. This is because preliminary
tests showed that using SBR, a significant fraction of the gas content
of the galaxy immediately became unbound and was lost, even at low
rotation speeds. The CR profile we use can be argued to have problems on
small radii. The constant value of the velocity means that the angular
velocity rises quickly when approaching the rotation axis. But here our
preliminary tests show that in practice this is not a problem. Because
the gas particles are placed randomly and we are working with floating
point numbers, no particle in practice will ever lie exactly on the
rotation axis. The gas particles in the central region initially
have a very high angular velocity, but interactions between the gas
particles quickly slow them down and convert the excess velocity to
heat, wich is then very quickly lost through the very efficient
radiative cooling. Also, since the gas is initially distributed
homogeneously over a large sphere, there is no large gas mass in this
central region. We did a number of test simulations with a variety of
alternative rotation profiles with more well-behaved central velocities:
\begin{itemize}
\item the rotation velocity rises as an arctangens with radius,
      with the asymptotic velocity equal to the constant value of the
      fiducial models.
\item the rotation velocity rises linearly out to 0.5~kpc and is then
      kept constant at the constant value of the fiducial models
      (``combined linear/constant'').
\item the rotation velocity rises as an arctangens with radius out to
      0.5~kpc and is then kept constant at the constant value of the
      fiducial models (``combined arctangens/constant'').
\end{itemize}
These simulations show no noticeable influence of the central velocity
profile on the behaviour of the models, see Fig. \ref{rotprof_check}.

We will henceforth refer to the set of CR models as the ``rotating
models''.

\subsubsection{Adding initial flattening}

As a flattening parameter we consistently use the axis ratio $q=c/a$ in
this paper, where $c$ is the shortened $z$ axis and $a$ is the axis in
the $x-y$ plane (we adopt axially symmetric models). Here, we flatten
both the gas and DM distributions. The parameters of the Kuz'min Kutuzov
profile naturally allow for introducing a flattening to the DM halo in
the form of an axial ratio \citep{kuzmin}, so the DM part of the
flattening is trivial. Flattening the homogeneous gassphere takes a
little bit more care. We need to scale up the axis in the $x-y$ plane
appropriately while scaling down the $z$ axis to ensure that the density
of the gas cloud remains the same. This is important because we do not
want to change any aspect of the dwarf galaxy models from
\citet{sander:dgmodels}, other than the geometry, and the initial
density is a very important one. Starting from a sphere with radius $r$,
if we want to achieve a flattening $q$ we produce it by calculating the
$a$ and $c$ axes as follows:
\begin{description}
 \item $a = rq^{-1/3}$
 \item $c = rq^{2/3}$
\end{description}
This will ensure that the volume of the ellipsoid is constant for any
value of $q$, and therefore so is the density.

In the remainder, we will refer to this set of models as the
``flattened models'' (although of course the rotating models also
become flattened eventually).

\subsection{Production runs}

In Table \ref{simulations} we show an overview of the run numbers and
specifications of our production runs. We simulate dwarf galaxies with a
range of masses, flattenings and rotation speeds using the dwarf galaxy
models of \cite{sander:dgmodels} from Table \ref{DG_models} and
additional methods for setting up the initial conditions as described
above. Both the gas and the dark matter components were represented by
200000 particles, and the simulations were evolved in time during 11.7
gigayears, corresponding to the time from $z=4.3$ to the present. 
The seeds used for sampling the particles from the specified density
profiles (see section \ref{section:basic_models}) are chosen at random
for all production runs.

\begin{table}

 \caption{ Grid of the production runs, given with runnumbers and
 specifications of the three used parameters: mass (first column, see
 Table \ref{DG_models}), initial flattening ($q$, last column) and
 initial rotation speed ($v_{i}$, 3 different rotation speeds) (see
 above).}

 \label{simulations}

   \begin{tabular}{l@{\extracolsep{4ex}}ccc@{\extracolsep{7ex}}r}
    \hline
    DG model &&&& $q$ \\
    \hline
    \hline
     & 0 \unit{km/s} & 1 \unit{km/s} & 5 \unit{km/s} & \\
    \hline
    \\
    C01 & 201 & 211 & 221 & 1 \\
        & 231 & 241 & 251 & 0.5 \\
        & 261 & 271 & 281 & 0.1 \\
    \\
    C03 & 203 & 213 & 223 & 1 \\
        & 233 & 243 & 253 & 0.5 \\
        & 263 & 273 & 283 & 0.1 \\
    \\
    C05 & 205 & 215 & 225 & 1 \\
        & 235 & 245 & 255 & 0.5 \\
        & 265 & 275 & 285 & 0.1 \\
    \\
    C07 & 207 & 217 & 227 & 1 \\
        & 237 & 247 & 257 & 0.5 \\
        & 267 & 277 & 287 & 0.1 \\
    \\
    C09 & 209 & 219 & 229 & 1 \\
        & 239 & 249 & 259 & 0.5 \\
        & 269 & 279 & 289 & 0.1 \\
    \\
    \hline

  \end{tabular}

\end{table}

\subsection{Preliminary evaluation of simulations}

Table \ref{simulations_summary} lists many different physical quantities
for all of the simulated dwarf galaxies in our set from Table
\ref{simulations}. These are the final values for these quantities,
evaluated at the end of the simulation, except those explicitly indexed
with `i', which are initial values. Broad-band colours are calculated
(with bilinear interpolation) using the models of \citet{vazdekis96},
who provide mass/luminosity values for SSPs according to metallicity and
age.  For those simulations that form little or no stars, making an
accurate evaluation of the physical parameters impossible, we simply
enter a ``$-$'' in Table \ref{simulations_summary}.

To evaluate our methods for setting up the initial conditions, we
discuss the C05 models below (see Table \ref{simulations}).

\begin{table*}
\begin{minipage}{180mm}

 \caption{Details of simulations. All physical quantities are evaluated
 at the end of the simulation (11.7 Gyr), except those indexed with `i',
 which are evaluated at the beginning. Columns: (1) model number (see
 Table \ref{DG_models}), (2) simulation number, (3) initial flattening
 (gas/DM), (4) initial rotation speed of gas [km/s], (5) spin parameter
 of gas in IC, (6) final gas mass [$10^{6} \mathrm{M_{\odot}}$], (7)
 stellar mass [$10^{6} \mathrm{M_{\odot}}$], (8) half-light radius
 [kpc], (9)(10) B-band and V-band magnitude, (11)(12) fitted S\'ersic
 parameters of surface brightness profile, (13) central stellar velocity
 dispersion along line of sight (edge-on) [km/s], (14)(15)
 luminosity-weighted metallicity (B-band), (16) final flattening of the
 stellar component (averaged over last 3 Gyr), (17) final stellar peak
 rotation speed [km/s]. Omitted values were irrelevant due to low
 stellar mass.}

 \label{simulations_summary}

\begin{tabular}{l*{16}{c}}
\hline                                
$\mathrm{model}$ & $\mathrm{run}$ & $q_{i}$ & $v_{i}$ & $\lambda$ & $M_{g,f}$ & $M_{\star}$ & $R_{e}$ & $M_{B}$ & $M_{V}$ & $I_{0}$ & $n$ & $\sigma_{1D,c}$ & $Z(\mathrm{Z_{\odot}})$ & $[Fe/H]$ & $q_{f}$ & $v_{f}$\\ 
\hline
 \\ 
C01 & 201 & 1 & 0 &  0.0 & 43.5 & 0.485 & 0.18 & -7.84 & -8.44 & 26.3 & 0.81 & 8.9 & 0.00036 & -1.907 & 0.99 & 0.7 \\ 
 & 211 & 1 & 1 & 0.007 & 43.5 & 0.519 & 0.15 & -8.23 & -8.76 & 26.5 & 0.62 & 8.2 & 0.00169 & -1.088 & 1.0 & 3.0 \\ 
 & 221 & 1 & 5 & 0.036 & 43.8 & 0.235 & 0.13 & -7.89 & -8.39 & 27.0 & 0.35 & 6.5 & 0.00247 & -1.005 & 0.96 & 2.9 \\ 
 \cline{2-17}  
 & 231 & 0.5 & 0 &  0.0 & 43.6 & 0.373 & 0.17 & -7.83 & -8.37 & 26.4 & 0.8 & 7.7 & 0.00113 & -1.241 & 0.74 & 1.0 \\ 
 & 241 & 0.5 & 1 & 0.009 & 43.6 & 0.419 & 0.17 & -7.91 & -8.51 & 26.3 & 0.78 & 7.3 & 0.0016 & -1.109 & 0.72 & 5.8 \\ 
 & 251 & 0.5 & 5 & 0.046 & 43.9 & 0.138 & 0.13 & -7.65 & -8.08 & 26.5 & 0.57 & 5.8 & 0.00188 & -1.106 & 0.69 & 1.9 \\ 
 \cline{2-17}  
 & 261 & 0.1 & 0 &  0.0 & 44.0 & 0.008 & 0.11 & -3.57 & -4.19 &  -- &  -- & 5.6 & 0.00073 & -1.561 &  -- & -0.1 \\ 
 & 271 & 0.1 & 1 & 0.014 & 44.0 & 0.004 & 0.11 & -2.65 & -3.25 &  -- &  -- & 4.6 & 0.00021 & -2.03 &  -- &  --  \\ 
 & 281 & 0.1 & 5 & 0.071 & 44.0 & 0.0 &  -- &  -- &  -- &  -- &  -- &  -- &  -- &  -- &  -- &  -- \\  \\ 
C03 & 203 & 1 & 0 &  0.0 & 67.6 & 2.316 & 0.22 & -9.8 & -10.34 & 25.2 & 0.68 & 12.1 & 0.00187 & -1.103 & 1.02 & 1.4 \\ 
 & 213 & 1 & 1 & 0.008 & 67.9 & 2.101 & 0.22 & -9.77 & -10.3 & 25.3 & 0.68 & 12.0 & 0.00219 & -1.058 & 0.93 & 4.9 \\ 
 & 223 & 1 & 5 & 0.039 & 68.3 & 1.671 & 0.25 & -9.84 & -10.36 & 25.2 & 0.59 & 10.7 & 0.00368 & -0.825 & 0.76 & 13.7 \\ 
 \cline{2-17}  
 & 233 & 0.5 & 0 &  0.0 & 67.7 & 2.283 & 0.25 & -9.65 & -10.25 & 25.9 & 0.55 & 11.3 & 0.00163 & -1.095 & 0.75 & 1.0 \\ 
 & 243 & 0.5 & 1 & 0.01 & 67.6 & 2.354 & 0.28 & -9.94 & -10.49 & 24.8 & 0.72 & 11.2 & 0.00371 & -0.795 & 0.7 & 8.2 \\ 
 & 253 & 0.5 & 5 & 0.05 & 68.6 & 1.377 & 0.26 & -9.69 & -10.21 & 25.4 & 0.56 & 10.0 & 0.00373 & -0.826 & 0.61 & 12.9 \\ 
 \cline{2-17}  
 & 263 & 0.1 & 0 &  0.0 & 69.3 & 0.691 & 0.19 & -8.88 & -9.39 & 24.5 & 0.92 & 9.2 & 0.00256 & -0.964 & 0.64 & 0.7 \\ 
 & 273 & 0.1 & 1 & 0.015 & 69.3 & 0.68 & 0.19 & -8.75 & -9.3 & 24.7 & 0.98 & 9.5 & 0.00262 & -0.97 & 0.6 & 4.1 \\ 
 & 283 & 0.1 & 5 & 0.076 & 69.7 & 0.293 & 0.15 & -8.32 & -8.78 & 24.8 & 0.85 & 8.3 & 0.00284 & -0.958 & 0.57 & 3.0\\  \\ 
C05 & 205 & 1 & 0 &  0.0 & 122.2 & 17.538 & 0.39 & -11.84 & -12.48 & 24.2 & 0.62 & 19.5 & 0.00445 & -0.699 & 1.0 & 0.4 \\ 
 & 215 & 1 & 1 & 0.009 & 114.1 & 25.561 & 0.45 & -12.1 & -12.79 & 24.4 & 0.59 & 20.9 & 0.00513 & -0.674 & 0.94 & 9.3 \\ 
 & 225 & 1 & 5 & 0.043 & 123.3 & 16.562 & 0.63 & -12.37 & -12.86 & 24.5 & 0.5 & 14.4 & 0.00623 & -0.61 & 0.53 & 24.2 \\ 
 \cline{2-17}  
 & 235 & 0.5 & 0 &  0.0 & 128.3 & 11.496 & 0.35 & -11.54 & -12.12 & 24.1 & 0.66 & 16.1 & 0.00291 & -0.886 & 1.03 & 1.2 \\ 
 & 245 & 0.5 & 1 & 0.011 & 111.2 & 28.514 & 0.59 & -12.76 & -13.3 & 24.5 & 0.36 & 20.4 & 0.00822 & -0.484 & 0.72 & 14.4 \\ 
 & 255 & 0.5 & 5 & 0.056 & 126.4 & 13.42 & 0.65 & -12.14 & -12.65 & 25.2 & 0.38 & 12.6 & 0.0063 & -0.609 & 0.45 & 24.2 \\ 
 \cline{2-17}  
 & 265 & 0.1 & 0 &  0.0 & 133.4 & 6.491 & 0.38 & -11.18 & -11.71 & 24.9 & 0.47 & 16.4 & 0.00305 & -0.871 & 0.67 & -0.3 \\ 
 & 275 & 0.1 & 1 & 0.017 & 129.5 & 10.351 & 0.47 & -11.87 & -12.38 & 24.8 & 0.43 & 16.5 & 0.0052 & -0.69 & 0.63 & 8.7 \\ 
 & 285 & 0.1 & 5 & 0.086 & 136.3 & 3.633 & 0.42 & -10.96 & -11.43 & 25.3 & 0.45 & 15.2 & 0.00436 & -0.771 & 0.49 & 11.7\\  \\ 
C07 & 207 & 1 & 0 &  0.0 & 84.9 & 175.6 & 0.55 & -14.21 & -14.92 & 22.8 & 0.67 & 32.8 & 0.01396 & -0.241 & 1.0 & 4.4 \\ 
 & 217 & 1 & 1 & 0.01 & 88.1 & 172.35 & 0.57 & -14.16 & -14.86 & 22.6 & 0.71 & 31.0 & 0.01525 & -0.191 & 0.88 & 18.6 \\ 
 & 227 & 1 & 5 & 0.048 & 173.6 & 87.62 & 1.09 & -14.0 & -14.53 & 24.6 & 0.39 & 18.5 & 0.00836 & -0.478 & 0.45 & 33.6 \\ 
 \cline{2-17}  
 & 237 & 0.5 & 0 &  0.0 & 144.9 & 115.46 & 0.7 & -14.09 & -14.7 & 23.5 & 0.48 & 32.1 & 0.01103 & -0.355 & 0.89 & 2.4 \\ 
 & 247 & 0.5 & 1 & 0.012 & 110.1 & 150.53 & 0.69 & -14.27 & -14.9 & 23.2 & 0.58 & 29.0 & 0.01346 & -0.257 & 0.73 & 18.5 \\ 
 & 257 & 0.5 & 5 & 0.062 & 197.8 & 63.68 & 1.17 & -13.51 & -14.09 & 25.2 & 0.38 & 16.5 & 0.00631 & -0.609 & 0.38 & 34.4 \\ 
 \cline{2-17}  
 & 267 & 0.1 & 0 &  0.0 & 195.6 & 65.82 & 0.81 & -13.95 & -14.45 & 24.1 & 0.41 & 26.6 & 0.00853 & -0.482 & 0.76 & 1.6 \\ 
 & 277 & 0.1 & 1 & 0.019 & 194.0 & 67.354 & 0.82 & -13.98 & -14.46 & 24.2 & 0.36 & 28.1 & 0.00892 & -0.455 & 0.68 & 8.5 \\ 
 & 287 & 0.1 & 5 & 0.095 & 240.6 & 21.168 & 0.84 & -12.91 & -13.37 & 25.0 & 0.38 & 18.2 & 0.00595 & -0.642 & 0.39 & 25.3\\  \\ 
C09 & 209 & 1 & 0 &  0.0 & 44.0 & 475.84 & 0.33 & -14.79 & -15.56 & 20.4 & 1.29 & 43.0 & 0.01594 & -0.109 & 1.02 & 2.7 \\ 
 & 219 & 1 & 1 & 0.011 & 46.7 & 473.57 & 0.48 & -14.8 & -15.58 & 21.0 & 1.11 & 40.3 & 0.01591 & -0.135 & 0.83 & 28.4 \\ 
 & 229 & 1 & 5 & 0.054 & 177.0 & 344.39 & 1.34 & -15.12 & -15.71 & 23.8 & 0.51 & 24.1 & 0.01363 & -0.238 & 0.4 & 45.3 \\ 
 \cline{2-17}  
 & 239 & 0.5 & 0 &  0.0 & 61.5 & 458.48 & 0.43 & -14.84 & -15.6 & 20.9 & 1.12 & 42.7 & 0.0167 & -0.104 & 0.83 & 2.0 \\ 
 & 249 & 0.5 & 1 & 0.014 & 67.8 & 452.32 & 0.58 & -14.86 & -15.63 & 21.8 & 0.89 & 37.3 & 0.01634 & -0.129 & 0.69 & 29.6 \\ 
 & 259 & 0.5 & 5 & 0.07 & 238.4 & 283.58 & 1.52 & -14.9 & -15.5 & 24.1 & 0.61 & 21.2 & 0.01078 & -0.348 & 0.36 & 44.5 \\ 
 \cline{2-17}  
 & 269 & 0.1 & 0 &  0.0 & 211.0 & 310.25 & 0.61 & -15.28 & -15.82 & 21.6 & 0.75 & 36.0 & 0.01835 & -0.086 & 0.79 & 0.4 \\ 
 & 279 & 0.1 & 1 & 0.021 & 225.5 & 295.97 & 0.74 & -15.18 & -15.74 & 22.5 & 0.58 & 35.6 & 0.01663 & -0.137 & 0.72 & 14.2 \\ 
 & 289 & 0.1 & 5 & 0.107 & 410.0 & 113.12 & 1.19 & -14.36 & -14.89 & 23.7 & 0.9 & 27.2 & 0.00812 & -0.491 & 0.39 & 29.7\\  \\ 
\hline 
\end{tabular}

\end{minipage}
\end{table*}

\subsubsection{Variance}

\label{variance_eval}

\begin{figure}
\centering \includegraphics[width=0.45\textwidth]{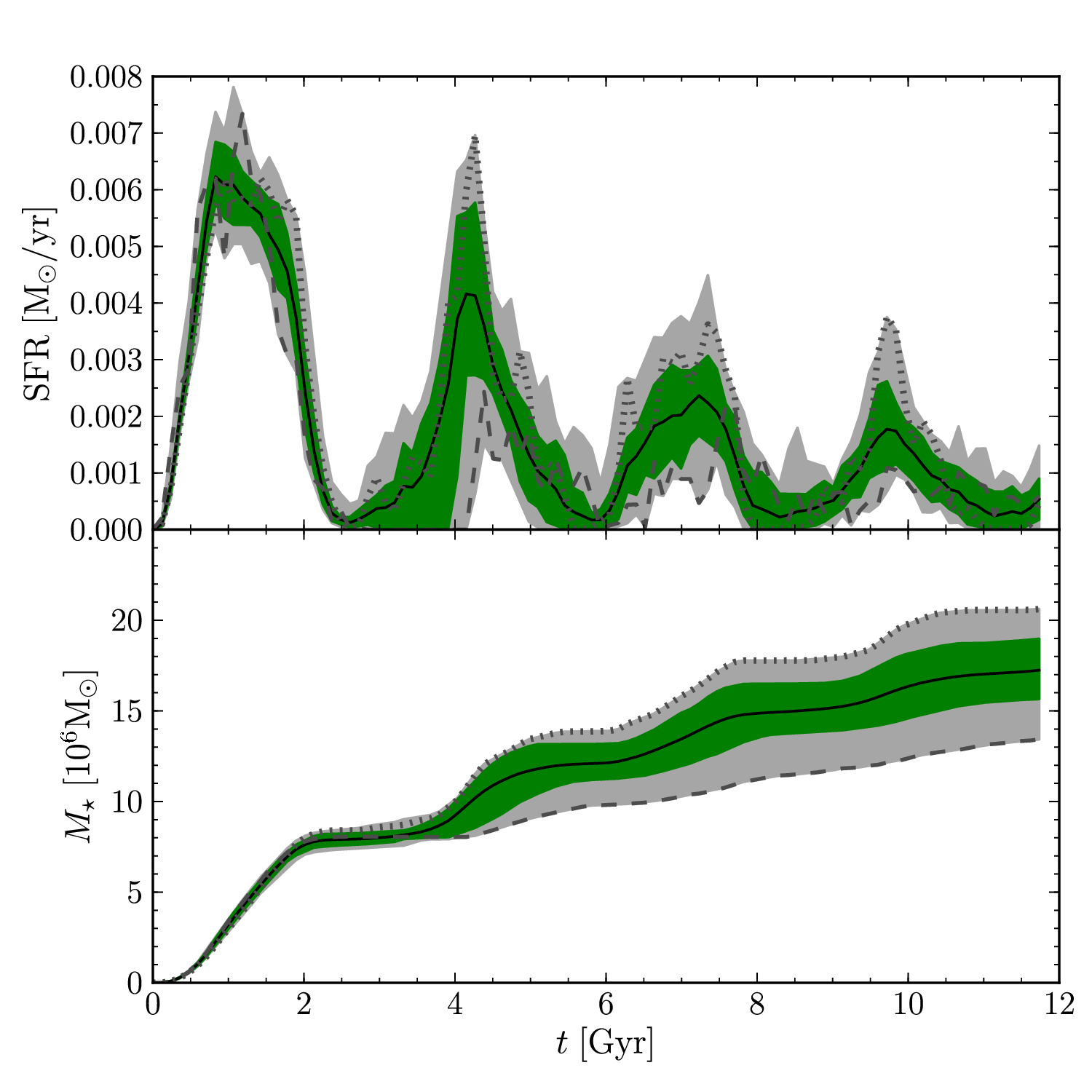}
 \caption{Depiction of the variance inherent to our models, using a
 set of 25 differently sampled initial conditions. {\em Upper panel}
 shows the evolution of the star formation rate, {\em lower panel} shows
 the evolution of the total stellar mass. The black line is the mean
 curve for our set. The green band shows the area between the 15.9th and
 84.1st percentile (which are linearly interpolated between the closest
 ranks, and would correspond to the $1\sigma$ interval if the underlying
 distribution was Gaussian), and the light grey band shows the area
 between the minimum and maximum value of our set. These percentiles and
 extrema are calculated in each time-bin (of which there are 100). The
 two dark grey lines show the evolution of two individual runs: the
 dashed and dotted line represent the runs which, respectively, produced
 the lowest and the highest total stellar mass at the end of the
 simulation.}

 \label{variance}

\end{figure}

We first make note of the inherent variance in our models. To this end
we have produced a set of 25 simulations of the basic spherical C05
model, with different samplings of the dark matter halo and the gas
sphere. For each simulation, different random seeds are used to
construct the initial condition. As shown by the 15.9th/84.1th
percentile area and the total range of the SFHs of this set of
simulations in Fig. \ref{variance}, the variance is significant,
allowing for a variety of SFHs. This is however not unexpected. Systems
of this kind, with stochastic star formation and feedback, are
inherently chaotic.  Small differences are continuously amplified and
can, over time, lead to large deviations. But on the other hand, more
importantly, the green band depicting the 15.9th/84.1th percentile area
(which would correspond to the $1\sigma$ interval if the underlying
distribution was Gaussian) shows quite clearly the generic behaviour of
the models. So we keep in mind that our models can show a spread in
their properties, but that they also exhibit a clear general behaviour.

\subsubsection{Rotating models}

\label{rotation_eval}

\begin{figure}
 \includegraphics[width=0.45\textwidth]{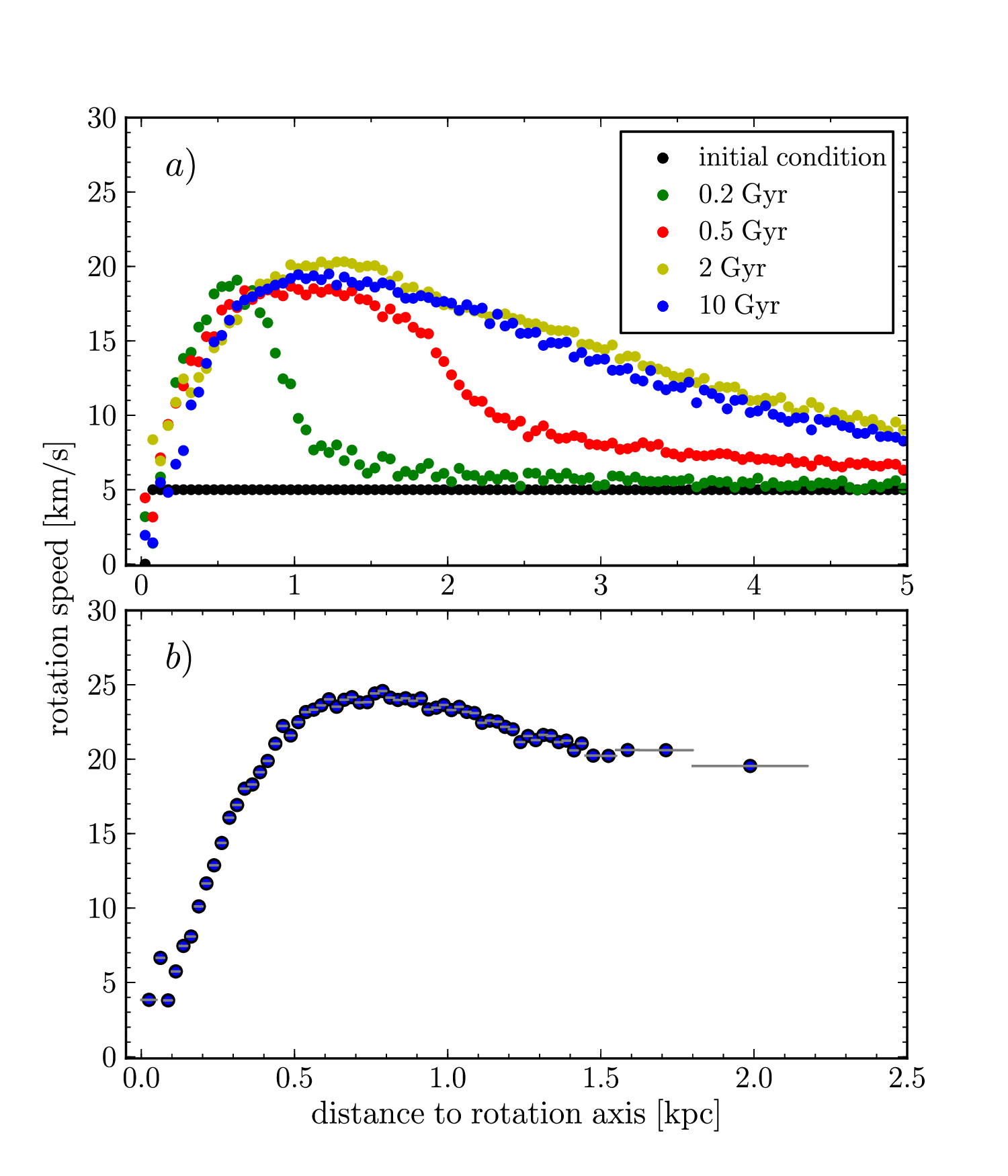}
 \caption{Rotation curves of our showcase model (225), {\em upper panel}
 displays the gas at different times during the simulation, {\em lower panel}
 displays the stars at 11.7 Gyr (with adaptive binning).}
 \label{rotation_curves}
\end{figure}

As described above, to obtain a rotating galaxy we add initial angular
momentum to the \textit{gas}. We need to check if this actually results
in a rotating \textit{stellar} component of the galaxy. In the upper
panel of Fig. \ref{rotation_curves}, we present the rotation curve of
the gas particles at different times. In the lower panel, we show the
final stellar rotation curve (at $11.7 \unit{Gyr}$). These are binned
profiles of tangential velocity versus distance to the $z$ axis, where
the profile value in every bin is the average rotation velocity per
particle in that bin.

The rotation profile of the gas rises due to the gas falling into the
potential well, and quickly evolves to a rather stable form, only
perturbed temporarily by the turbulence caused by strong star formation
events. This ``steady state'' is a consequence of the balance between
cooling, which makes the rotating gas sink inwards, and supernova
feedback, which heats and disperses gas. The stars that form from the
gas finally follow a rotation profile that rises out to one half-light
radius and flattens off beyond that radius. This confirms that using CR
initial conditions for the gas is adequate to achieve stable, rotating
dwarf galaxies. The final rotation speed of the stars which is included
in Table \ref{simulations_summary} is the peak value of this rotation
curve.

\subsubsection{Flattened models} \label{flattening_eval}

\begin{figure}
\centering

 \includegraphics[width=0.45\textwidth]{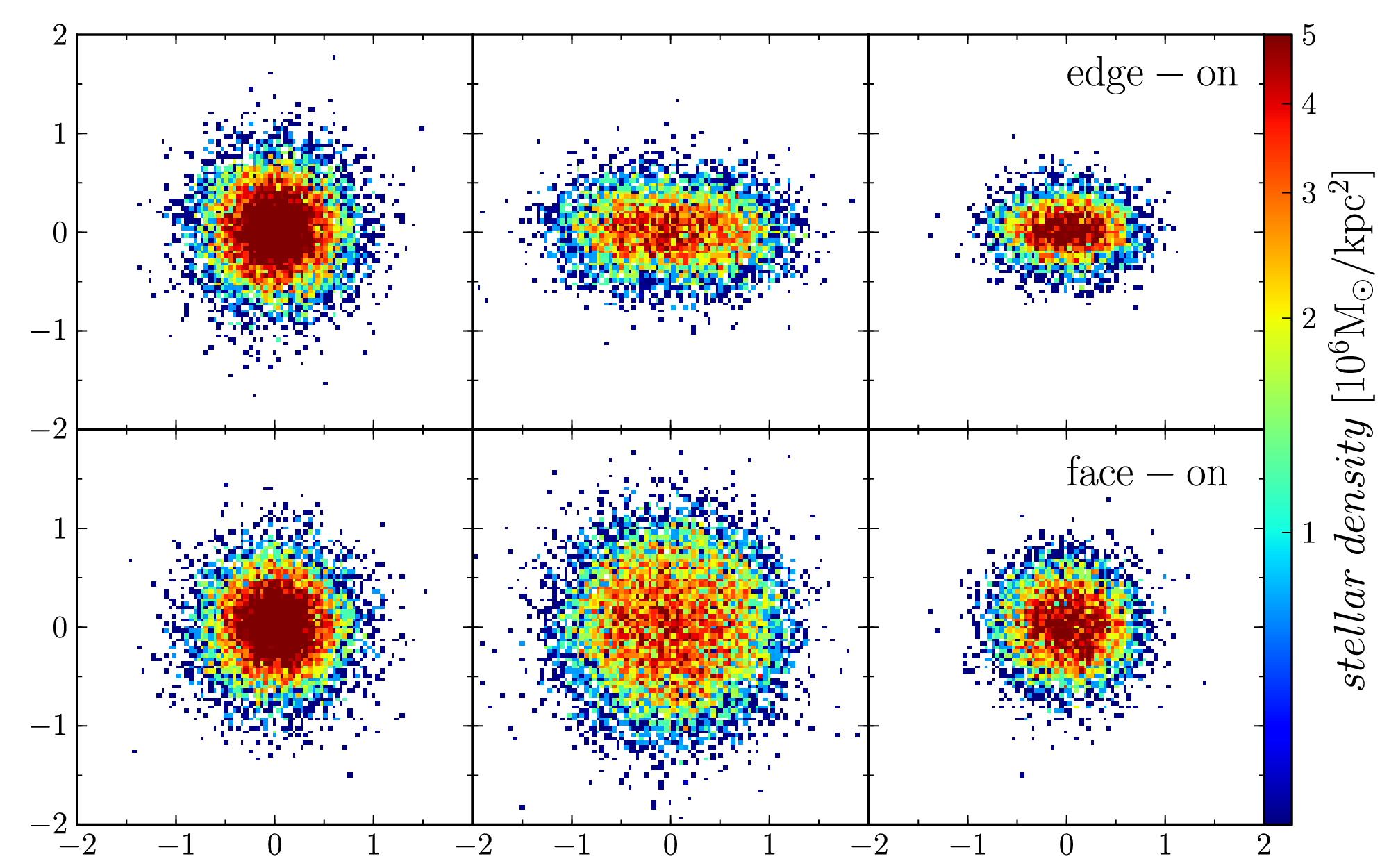}
 \caption{Edge-on and face-on views of the stellar distributions of a
 non-rotating spherical model (left, 205), a rotating model (middle,
 225) and a flattened, non-rotating model (right, 265). All are slices
 of thickness $0.4$ kpc, axes are in kpc, and color denotes projected
 stellar density.}

\label{flat_snapshots}

\end{figure}

\begin{figure}
 \includegraphics[width=0.48\textwidth]{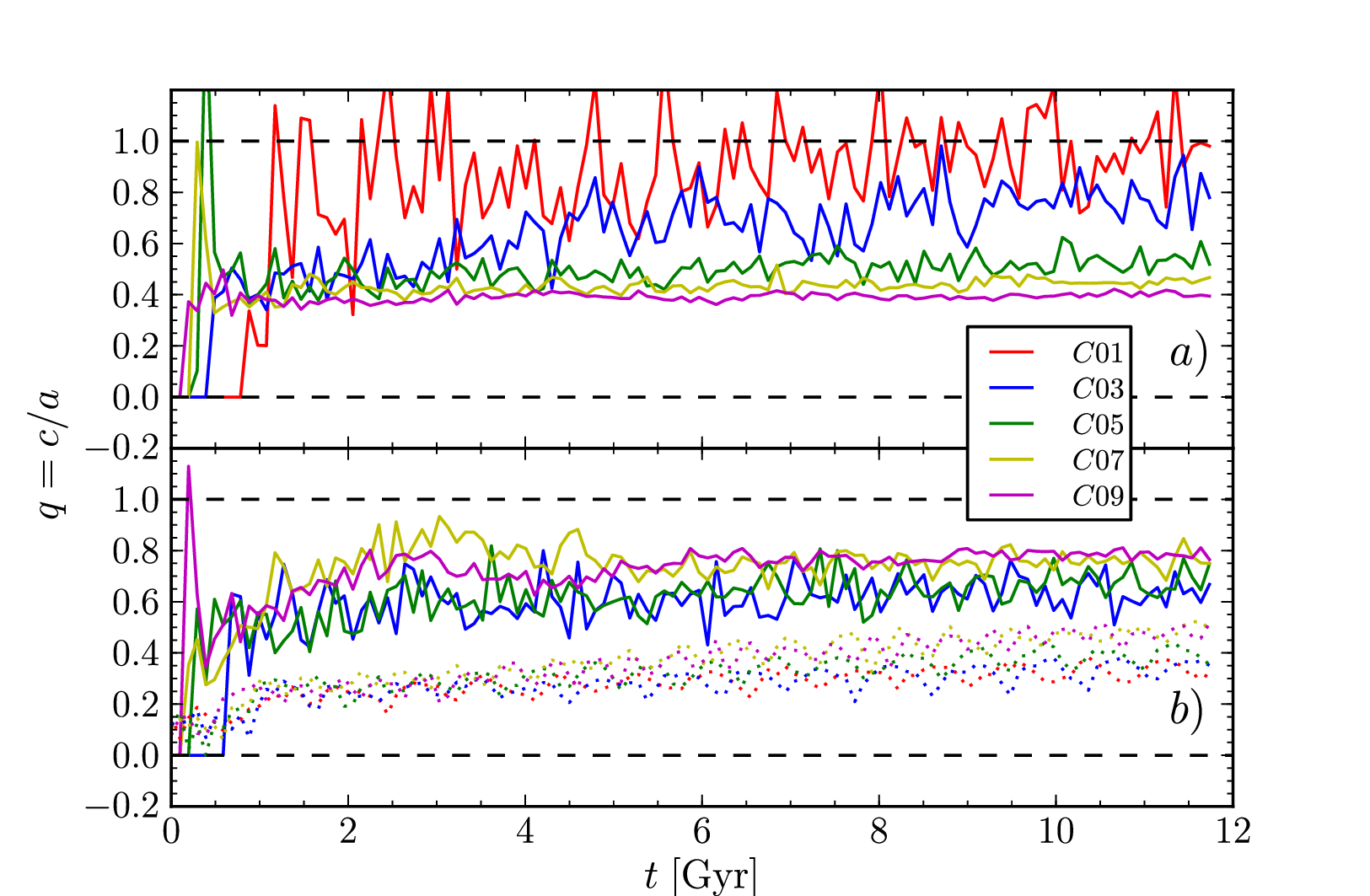}
 \caption{Evolution of the flattening parameter $q=c/a$ of the stellar
 component during the simulations for different galaxy masses. {\em
 Upper panel}:~rotating models (all with $v_{i}=5 \unit{km/s}$), {\em
 lower panel}:~flattened models (all with $q_{i}=0.1$). Per panel all
 properties are identical, except for the mass. Only galaxies with an
 appreciable stellar mass are shown (see Table
 \ref{simulations_summary}). Dotted lines show the $q$ of the DM
 component, only shown in the bottom panel because the DM consistently
 has $q=1$ in the top panel.}

 \label{flat_stability}

\end{figure}

In Fig. \ref{flat_snapshots} we show three of our model galaxies (a
non-rotating spherically symmetric model on the left, a rotating model
and a flattened model) to see the resulting flattenings at the end of
the simulations. The latter two both show a considerable and stable
flattening (see the C05 model in Fig. \ref{flat_stability}). Some trends
become apparent when looking at the total mass range in Table
\ref{simulations_summary} and Fig. \ref{flat_stability}. It appears that
the stability of $q$ for the stellar component significantly increases
with rising mass in the rotating models. Only in the least massive
models does $q$ rise significantly with time. The more massive models
all exhibit a stable flattening around $q \sim 0.4-0.5$, so our model
dwarf galaxies are relatively ``thick''. In non-rotating models with an
initially strongly flattened halo, the halo thickens and it turns out to
be impossible to make stellar bodies more flattened than $q \sim
0.6-0.8$. As can be read from Table \ref{simulations_summary}, combining
initial rotation and initial flattening helps somewhat to achieve
stronger flattenings in the least massive models.

\begin{figure}
 \includegraphics[width=0.48\textwidth]{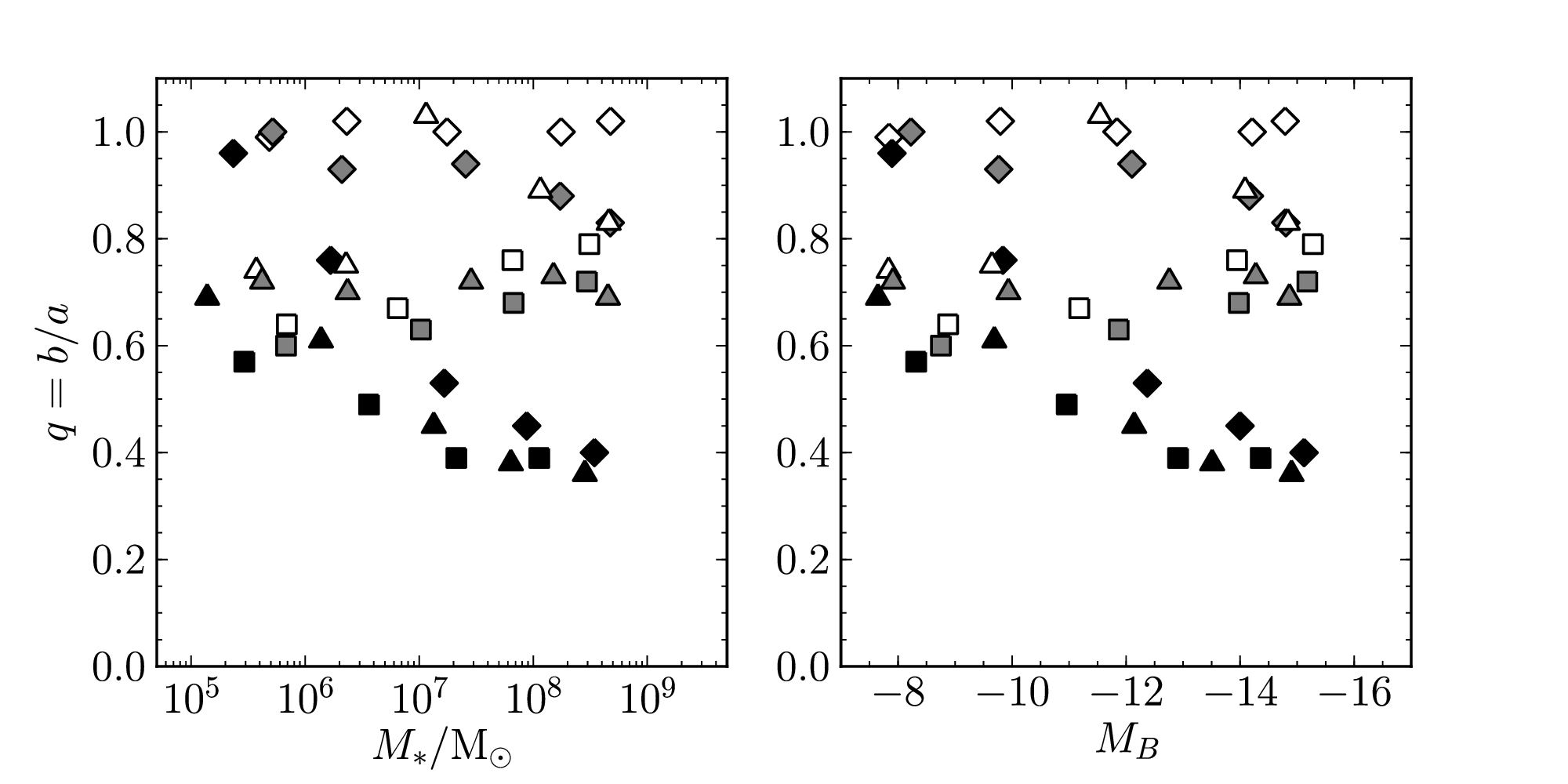}
 \caption{Axis ratios of all our models. {\em Left panel}: versus
 stellar mass (in $\mathrm{M_{\odot}}$), {\em right panel}: versus
 B-band luminosity. The different initial rotation speeds are indicated
 with color (white: $0$ km/s, grey: $1$ km/s, black: $5$ km/s), the
 initial flattenings are indicated with symbol shapes (lozenge: $1$,
 triangle: $0.5$, square: $0.1$).}

 \label{sanchez_plot}

\end{figure}

We thus note from Table \ref{simulations_summary} and Fig.
\ref{flat_stability} that we are not able to make extremely flat
galaxies. In the work of \citet{roychowdhury:flatdist} a collection of
dIrrs from the FIGGS survey is investigated, and they find from the
flattening distribution a mean axial ratio $\langle q \rangle \approx
0.6$ for the \textit{HI disks}. Similar values are obtained by
\citet{binggeli:flatdist}, \citet{staveley:flatdist},
\citet{hunter:flatdist}, \citet{sung:flatdist} and \citet{sanchez10} for
the \textit{stellar} content of dwarf galaxies. Also other simulations
suggest that low mass galaxies are not born as thin discs, but as thick,
puffy systems (e.g. \citealt{kaufmann07}; and to a lesser extent the
(more massive) models of \citealt{governato10}, which are still not
extremely flat). The reason for this is sought in the increasing
importance of turbulent motions, plausibly caused by star formation and
feedback, with respect to rotational motion in low mass systems
\citep{kaufmann07,roychowdhury:flatdist,sanchez10}.  Besides the
moderate value of the flattening itself, we also qualitatively reproduce
the trend with galaxy mass from \citet{sanchez10}. All our simulations
are below their `limiting mass' of $M_{*} \approx 2 \times
10^{9}~M_{\odot}$, and indeed for simulations with identical initial
setup, the final stellar bodies thicken with decreasing mass. This can
be seen in Fig. \ref{sanchez_plot}, where we mimic their Fig. 1 (the
leftmost and rightmost panels). Our most massive, flat models ($M_{*}
\approx 3.4 \times 10^{8}~M_{\odot},~q \approx 0.4$) connect nicely to
models DG1 and DG2 of \citet{governato10}, who are slightly more massive
and slightly flatter ($M_{*} \approx 4.8 \times 10^{8}~M_{\odot},~q
\approx 0.35$).

\subsubsection{Galaxy mass and concentration}

For a rotating model the half-light radius ($R_{eL}$), defined as the
radius of the sphere containing half of the light) is considerably
larger than that of a sperical model, as can be seen in Table
\ref{simulations_summary} and Fig. \ref{flat_snapshots}. The total
stellar mass usually decreases when adding significant
rotation. Non-rotating flattened models on the other hand are
generally not much larger than the spherical models, sometimes even
smaller. The half-light radius decreases slightly in flattened
galaxies at lower masses, and increases slightly at higher masses
(with respect to the spherical model). The total stellar mass
decreases with increasing flattening.

The rotating models are thus spatially more extended than their
spherical progenitor and at the same time they generally are also less
massive (in stellar mass) so they are considerably less centrally
concentrated. The flattenend, non-rotating models are usually less
spatially extended than the spherical models and also less massive, so
they have similar central concentrations (see also
Fig. \ref{flat_snapshots}).

\section{Analysis}
\label{section:analysis}

In this section we present a more extensive analysis of our production
runs. 
\subsection{Metallicity profiles}
\label{section_metprof}

\begin{figure*}
\begin{minipage}[t]{\columnwidth}
\begin{center}
\includegraphics[width=\textwidth,clip]{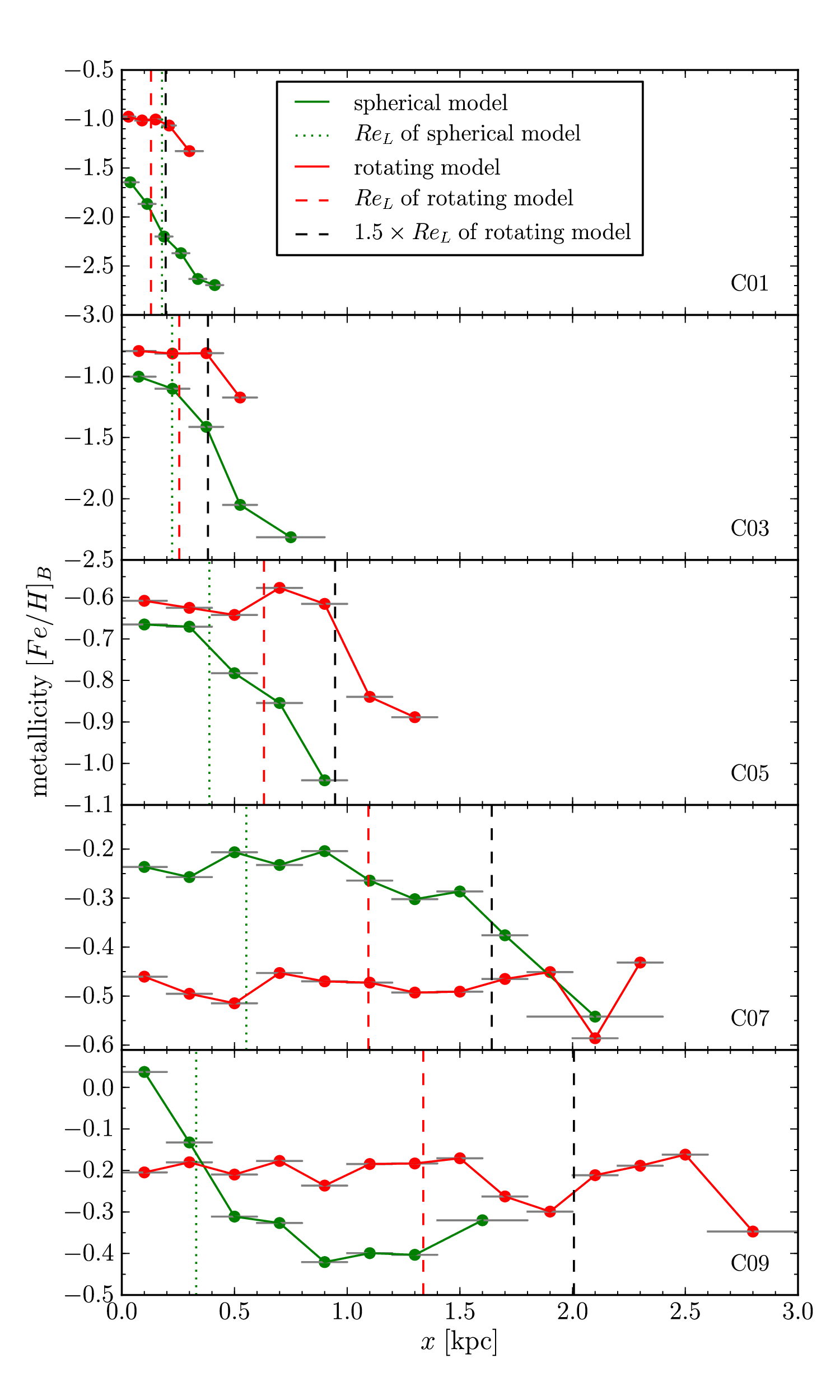}
\caption{Metallicity profiles of our productions runs, see Table
 \ref{simulations}. Each frame compares, for a certain galaxy mass, the
 spherical model (20$x$) with the fastest rotating model (22$x$). The
 $R_{eL}$ of each model is also indicated with a dashed line, and for
 the rotating model we also show $1.5 \times R_{eL}$ for indicative
 purposes. Adaptive binning was used to produce these profiles, the
 width of each bin being indicated by a horizontal grey bar.\label{metprof_rot}}
\end{center}
\end{minipage}
\hfill
\begin{minipage}[t]{\columnwidth}
\begin{center}
\includegraphics[width=\textwidth,clip]{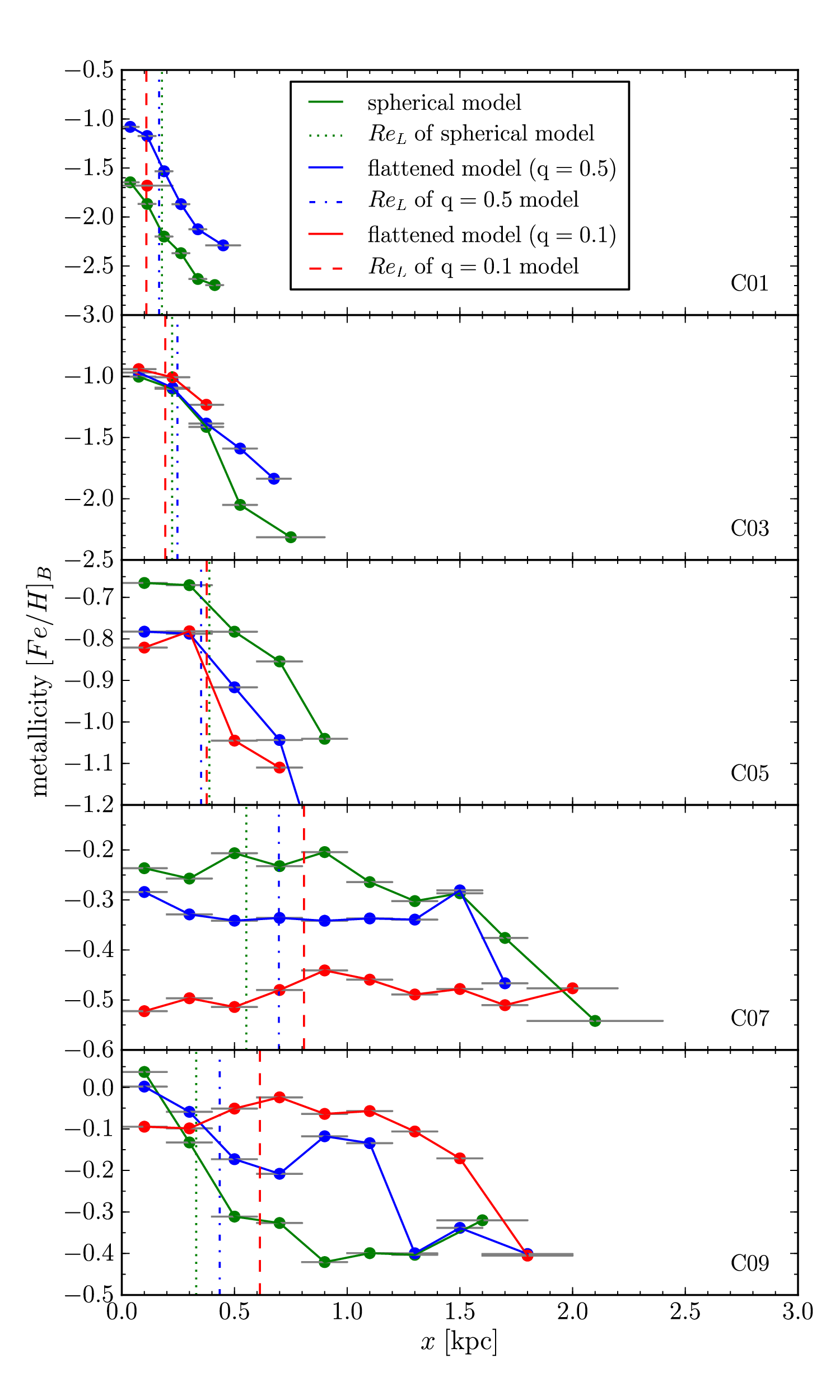}
\caption{Metallicity profiles of our productions runs, see Table
  \ref{simulations}. Each frame compares, for a certain galaxy mass,
  the spherical model (20$x$) with the 2 non-rotating flattened models
  (23$x$ and 26$x$). Further details of the plot are similar to those
  of Fig. \ref{metprof_rot}. \label{metprof_flat}}
\end{center}
\end{minipage}
\end{figure*} 

Looking at Fig. \ref{metprof_rot} we see some interesting results
concerning the metallicity profiles of the rotating galaxies. For a
range of galaxy masses we compare the metallicity profiles of the
spherical models with those of the fastest rotating models from Table
\ref{simulations} (with $v_{i}= 5 \unit{km/s}$) in Fig.
\ref{metprof_rot}. The metallicity profiles of the spherical models
almost always show a clear, negative gradient, while the profiles of the
rotating models are always significantly flatter. For a proper
comparison between different models of different sizes, the half-light
radius of each simulation is also indicated on the plots with a dashed
vertical line. We note that the rotating models can be considered to
have flat profiles out to 1.5 times $R_{eL}$, while the spherical models
usually show a fall-off well before that. Noticeably, the mean [Fe/H] of
the lower mass models appears to be significantly higher when rotating,
while the opposite is true for the higher mass models. This will be
discussed further on in section \ref{section_scaling_metallicity}.

Fig. \ref{metprof_flat} shows the same quantities for some of the
flattened models. Surprisingly, there appears to be no obvious trend
between the flattening and the shape of the metallicity profile, with
most galaxies showing strong negative metallicity gradients. We show
all models from Table \ref{simulations} which received an initial
flattening but no initial rotation, so for each mass we have 2
different degrees of initial flattening ($q=0.5$ and $q=0.1$). It is
clear that the flattening generally has no significant effect on the
metallicity gradient, almost all simulations have a negative
slope. Only in the most massive ones, or where the spherical model
does not have a strong gradient to begin with, does the initial
flattening appear to have some ability to somewhat flatten the
metallicity profile.

\subsection{Star formation histories}
\label{section_SFH}

\begin{figure*}
\begin{minipage}[t]{\columnwidth}
\begin{center}
\includegraphics[width=\textwidth,clip]{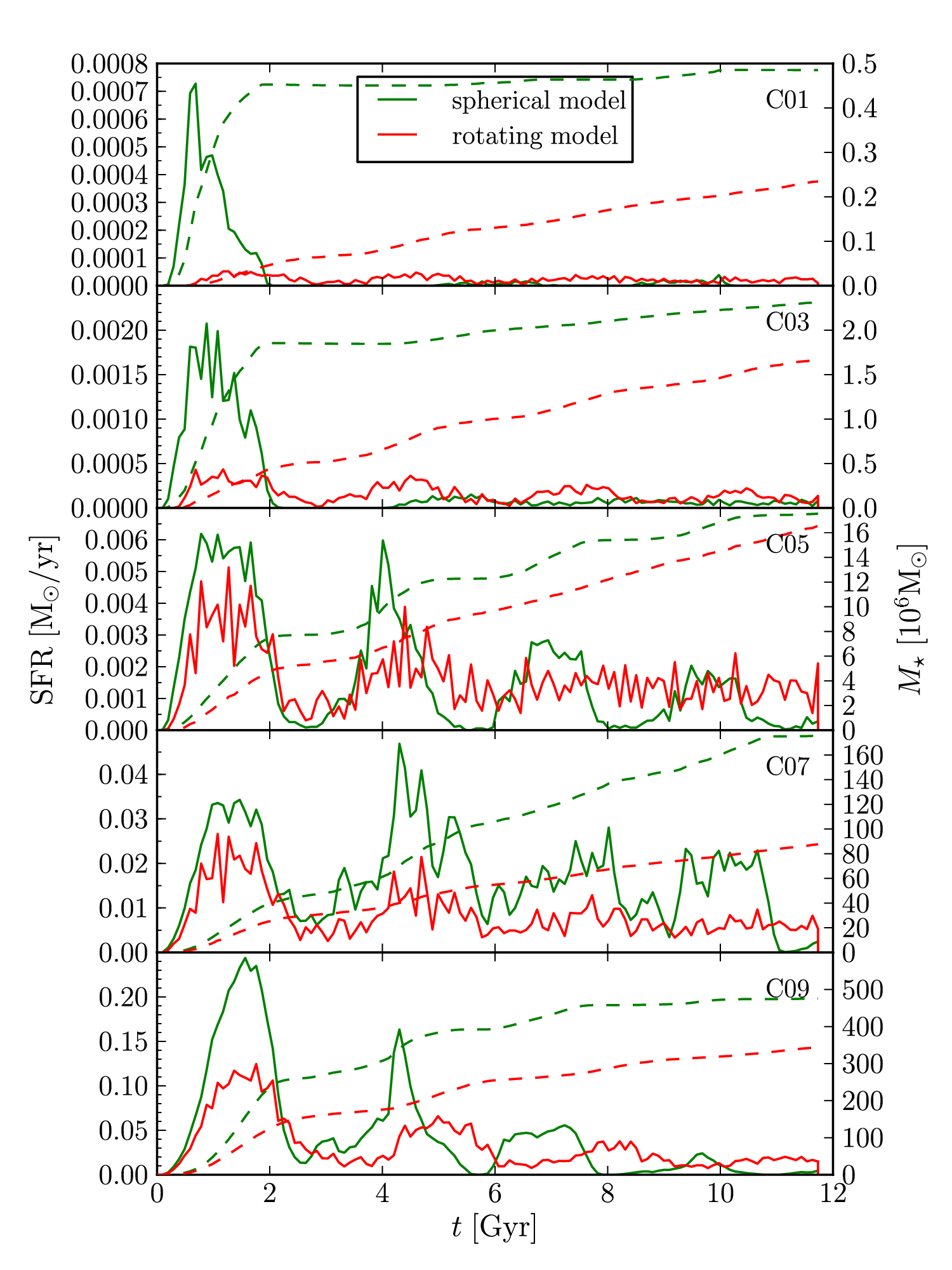}
\caption{SFHs of our productions runs, see Table \ref{simulations}. The
 same runs as in Fig. \ref{metprof_rot} are plotted here, comparing
 spherical and rotating models for different masses. Both their SFH
 (solid lines) and evolution of their stellar mass (dashed lines) are
 shown. \label{SFH_rot}}
\end{center}
\end{minipage}
\hfill
\begin{minipage}[t]{\columnwidth}
\begin{center}
\includegraphics[width=\textwidth,clip]{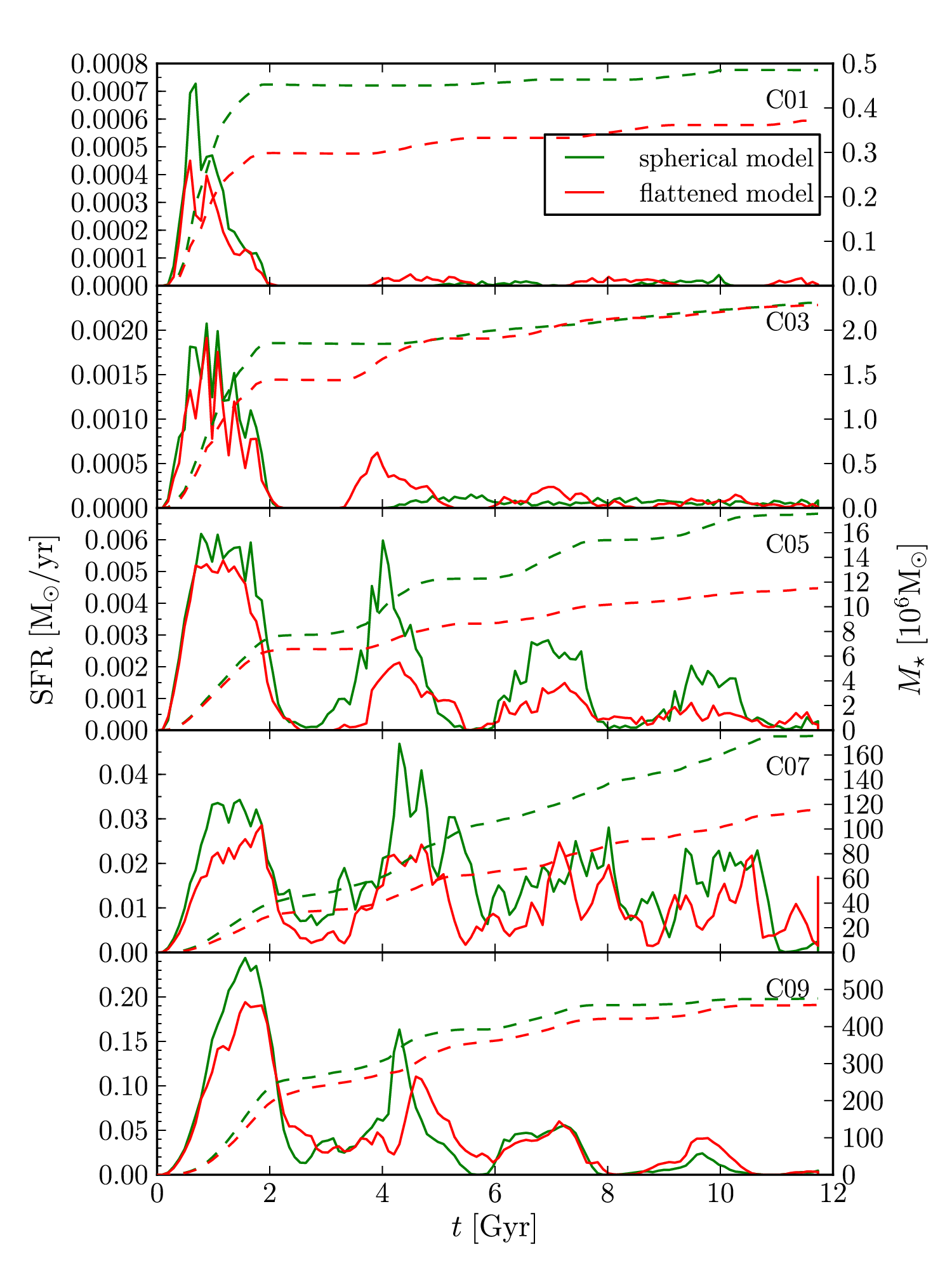}
\caption{SFHs of our productions runs, see Table \ref{simulations}. The
 runs with $q=0.5$ from Fig. \ref{metprof_flat} are plotted here,
 comparing spherical and flattened models for different masses. Similar
 to Fig. \ref{SFH_rot}. \label{SFH_flat}}
\end{center}
\end{minipage}
\end{figure*} 

Next we turn our attention to the star formation, and for this we look
at Fig. \ref{SFH_rot}, where star formation histories (SFH) of
different simulations are shown. Rotation also seems to have a
significant influence here. We again compare the spherical models with
the fastest rotating models from Table \ref{simulations} for a range
of galaxy masses in Fig. \ref{SFH_rot}, where we show the evolution
of the produced stellar mass expressed in solar mass per year
($\mathrm{M_{\odot}/yr}$). The total stellar mass of the galaxy is
plotted in dashed lines alongside the SFHs. 

Non-rotating spherical models typically have ``breathing'' or
``bursty'' SFHs, with strong SF peaks a few Gyr long, separated by
quiescent periods where the star formation rate (SFR) essentially goes
to zero \citep{sander:dgmodels, sti07, re09}. The strength and duration
of these peaks, as well as the intermittent pauses, depend mainly on
galaxy mass.

The models with rotation, however, are able to reduce this burstiness
and make the SFH much more continuous. Periods of increased star
formation still exist, alternated with lulls, but the SFR never drops
down to zero. The effectiveness of reducing the SF peaks varies in our
simulations, and depends primarily on the galaxy mass. In the least
massive models, which in the non-rotating spherical case show the most
extreme bursty behaviour (one big initial burst almost completely
shutting down further SF activity), the effect of adding rotation is
most noticeable. The SFR now becomes virtually flat. The more massive
models still show some SF fluctuations but not nearly as pronounced as
in their spherically symmetric analogs.

Flattening on the other hand does not have a large effect on the star
formation history of the galaxies. When looking at Fig. \ref{SFH_flat}
we can see that flattening, unlike rotation, generally does not induce
major qualitative differences in the SFH. The SFHs are generally very
much like the SFHs of the spherical models, still having large peaks
separated by periods with zero star formation.

\subsection{Gas structure}
\label{section_gas_structure}

\begin{figure*}
\begin{minipage}{180mm}
 \includegraphics[width=0.5\textwidth]{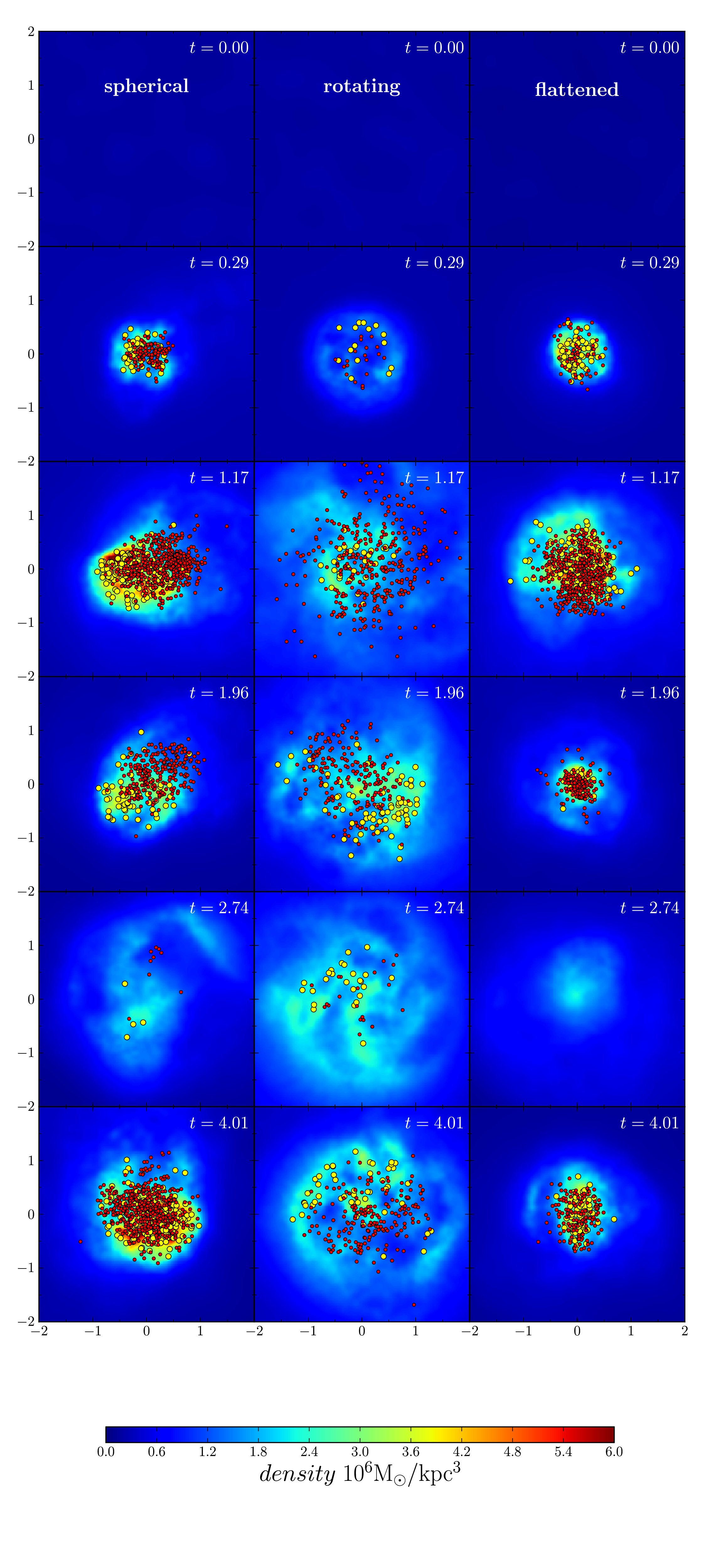}
 \includegraphics[width=0.5\textwidth]{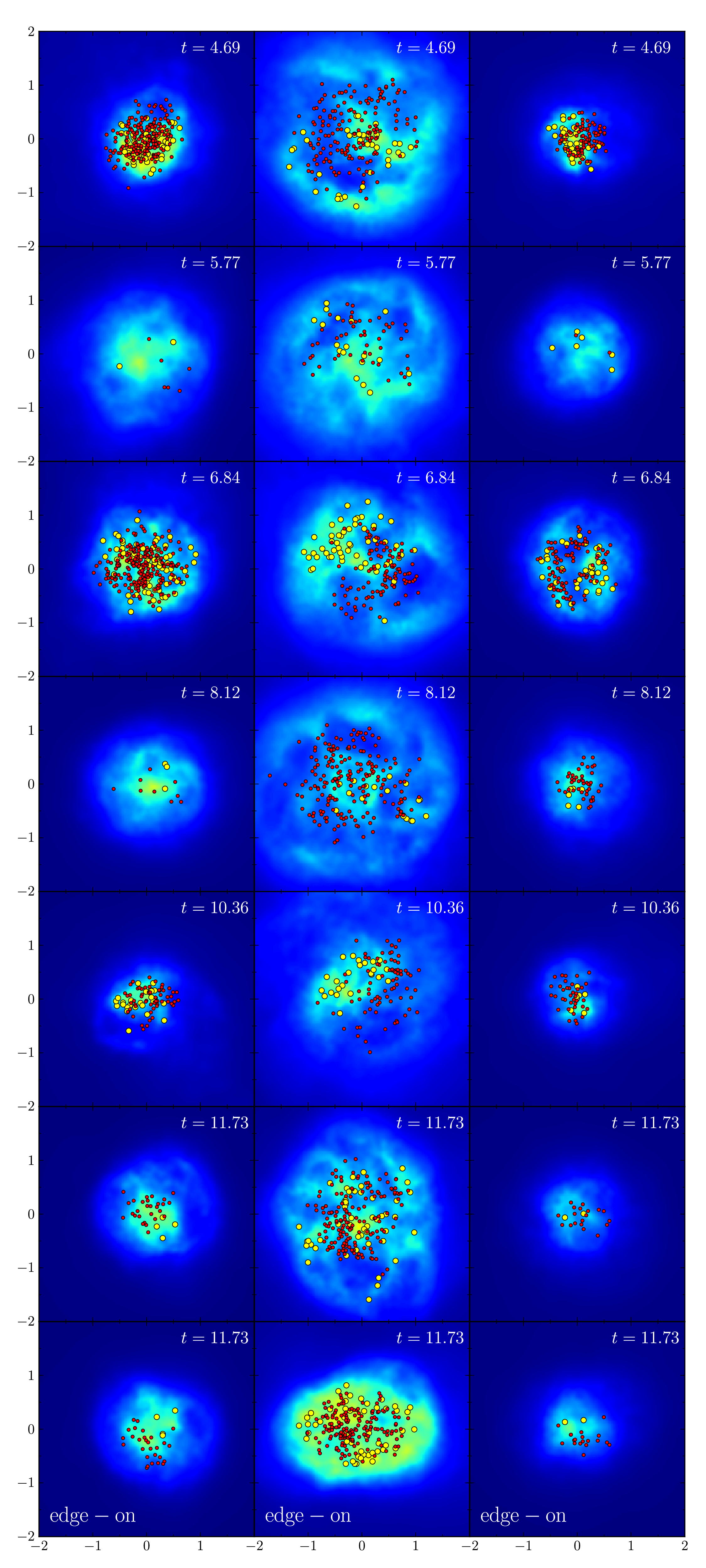}
 \caption{A series of snapshots from the evolution of 3 simulated dwarf
 galaxies based on the C05 DG model. {\em Left column}:~basic spherical
 model (205), {\em middle column}:~rotating model (225), {\em right
 column}:~flattened, non-rotating model (265). The bottom of the left 3
 columns continues on the top of the right 3 colums. Snapshots show
 rendered gas density (colorbar), {\em new} star particles (yellow dots,
 stellar age $<20$Myr) and {\em recent} star particles (red dots,
 $40$Myr $<$ stellar age $<100$Myr). The label in the top right corner
 indicates the time in the simulation (Gyr), and all galaxies are shown
 face-on in the $x-y$ plane, except the last 3 snapshots which are
 edge-on in the $x-z$ plane (axes are in kpc). A full, high quality
 animation can be found online, see text.  \label{snapshots}}
\end{minipage}
\end{figure*}

The structure of the gas of dwarf galaxies is another typical
characteristic that we will consider. The best observations available in
this respect come from the Magellanic Clouds; for instance \citet{smcHI}
present HI data for the LMC. Another very useful source of observational
data about the HI gas content and structure of dwarf galaxies is the
THINGS survey \citep[The HI Nearby Galaxy Survey;][]{walter08, weisz09}.
These studies show that the neutral hydrogen gas of dIrrs generally
shows an obvious ``bubble structure'', consisting of myriad spherical
low density regions or ``holes'' in the gas with a large range of sizes.

The origin of these holes has long been attributed to stellar feedback
by single-age new-born stellar clusters \citep[][and references
therein]{weaver77,mccray87}. However, for the LMC it has proven to be
not at all evident to correlate HI holes or shells with H$\alpha$
emission \citep{kim99,book08}. Holmberg II has similar issues, with
H$\alpha$ not tracing the holes, and the stellar ages found therein not
corresponding well with the kinematical age of the holes
\citep{stewart00, rhode99, weisz09}. Studying this last galaxy in
detail, \citet{weisz09} propose a multi-age model, where HI holes are
created by stellar feedback from multiple generations of star formation
spread out over tens to hundreds of Myr. This model is supported by the
fact that H$\alpha$ and 24~micron emission, which trace the most recent
SF, do not correlate well with HI holes, while UV emission, which traces
SF over roughly the last 100 Myr, correlates much better. The concept of
a single age for a hole is rendered ambiguous.

In Fig. \ref{snapshots} we show the structure of 3 of our simulated
dwarf galaxies, in a sequence of snapshots taken throughout their entire
evolution (all shown face on). The projected gas density is rendered as
the background color (see colorbar), and two different age selections of
the stellar population are plotted. In accordance with \citet{weisz09}
we choose these to represent the {\em newest} stars (yellow dots,
stellar age $<20$Myr) which would be detected in H$\alpha$, and the {\em
recent} stars (red dots, $40$Myr $<$ stellar age $<100$Myr) which would
show up in UV. The gap between the two populations serves to provide a
clearer distinction between them on the plots. These 3 simulations
compare a spherical model (205, left column), a flattened non-rotating
model (265, right column) and a rotating model (225, center column), all
based on the C05 model (see Tables \ref{DG_models} and
\ref{simulations}). The specific snapshot times have been selected to
represent ``interesting'' moments in the galaxies' SFHs, coinciding with
SF peaks or lulls, see Figs. \ref{SFH_rot} and \ref{SFH_flat}. A full,
high quality animation, corresponding to Fig. \ref{snapshots}, can be
found online\footnote{HD video:
http://www.youtube.com/watch?v=L2OWqfM1azo \\ YouTube channel of
Astronomy department at Ghent University:
http://www.youtube.com/user/AstroUGent \\ YouTube playlist with all
additional material for this paper:
http://www.youtube.com/user/AstroUGent\#g/c/EFAA5AAE5 C5E474D }.

\subsubsection{Spherical simulations}

In the spherical models gas collapses to the center and forms stars that
collectively blow out the gas through feedback, preferably to one side
in a so-called chimney (see snapshots at $t=1.17$). Over time, the gas
cools and re-collapses after which star formation can resume again
(snapshots at $t=2.74$, $t=5.77$ and $t=8.12$). This cycle continues
throughout the entire evolution. There is no significant difference in
the correlation with the local gas density between the two stellar
populations shown in Fig. \ref{snapshots}. Both populations are
centrally concentrated, and so is the gas density.

Overall there is little small-scale structure:~the behaviour of the gas
takes place on a large, collective scale. This becomes particularly
apparent when comparing to the rotating model, discussed further on in
\ref{section_structure_rotating}. An occasional small bubble can be
spotted in the gas when the galaxy is forming stars (e.g. snapshot at
$t=4.69$, on the upper side of the galaxy). This large-scale behaviour
translates into the characteristics of the SFHs of the spherical models
discussed before:~large SF peaks separated by quiescent periods.

\subsubsection{Flattened simulations}

The structure of the gas in the flattened dwarf galaxies is quite
similar to the spherical ones. Large-scale behaviour with a centralized
structure is still very much the case, which again can be connected to
the discussion and conclusions about the SFHs of the flattened
galaxies. Small-scale structure is not significantly more present than
in the spherical models, and the previous discussion of the evolution of
the spherical models is equally valid for the initially flattened,
non-rotating models.

\subsubsection{Rotating simulations}

\label{section_structure_rotating}

The structure of the gas content of the simulated dwarf galaxies is
noticeably different when adding rotation. There is now much more
small-scale structure in the gas. A ``bubble structure'' emerges in
the gas, caused by the stellar feedback of individual star particles
(snapshot at $t=0.29$) or small pockets of star particles (very clear
at e.g. $t=6.84$). Apparently, the influence of stellar feedback has
become more {\em local}, and the gas does not exhibit the same {\em
  global}, large scale behaviour seen in the rotationless
models. 

\begin{figure*}
\begin{minipage}[t]{\columnwidth}
\centering
 \includegraphics[width=\columnwidth]{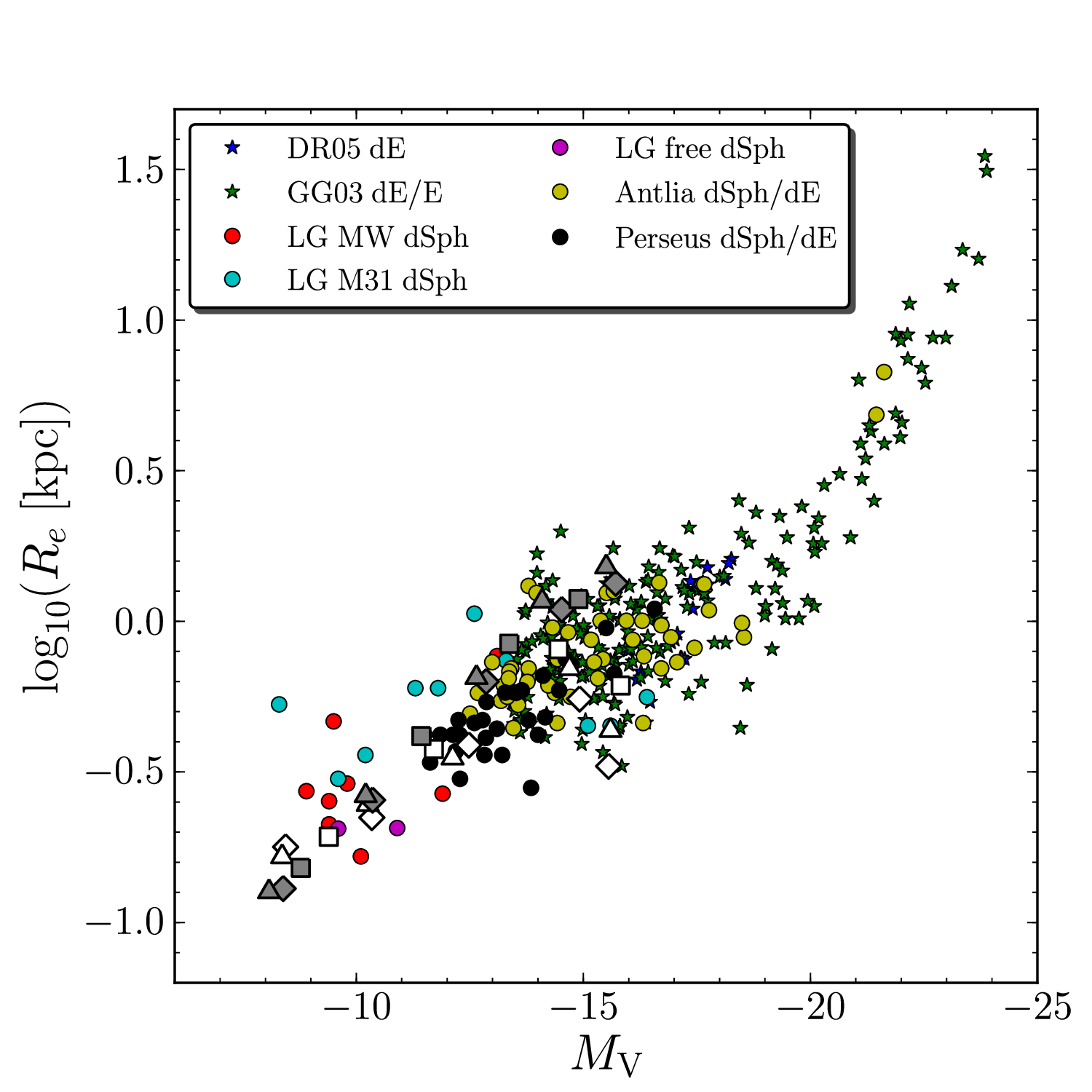}
 \caption{Half-light radius versus V-magnitude. White symbols are our
 non-rotating galaxies; the grey ones are the fastest rotating galaxies
 (see text). All other points on the plot are observational data found
 in \citet{rij09}: LG data come from
 \citet{peletier1993,irwin1995,saviane1996,gre03,mccon2006,mccon2007,zucker2007},
 Fornax data from \citet{mieske2007}, Antlia data from
 \citet{smith2008}, Perseus data from \citet{rij09}, other data from
 \citet{gra03} (GG03), \citet{rij05} (DR05). The symbol shapes
 distinguish the initial halo flattenings: simulations with $q=1$ are
 shown with lozenges, $q=0.5$ with triangles, $q=0.1$ with
 squares. \label{scaling_MV_re}}
\end{minipage}
\hfill
\begin{minipage}[t]{\columnwidth}
\centering
 \includegraphics[width=\columnwidth]{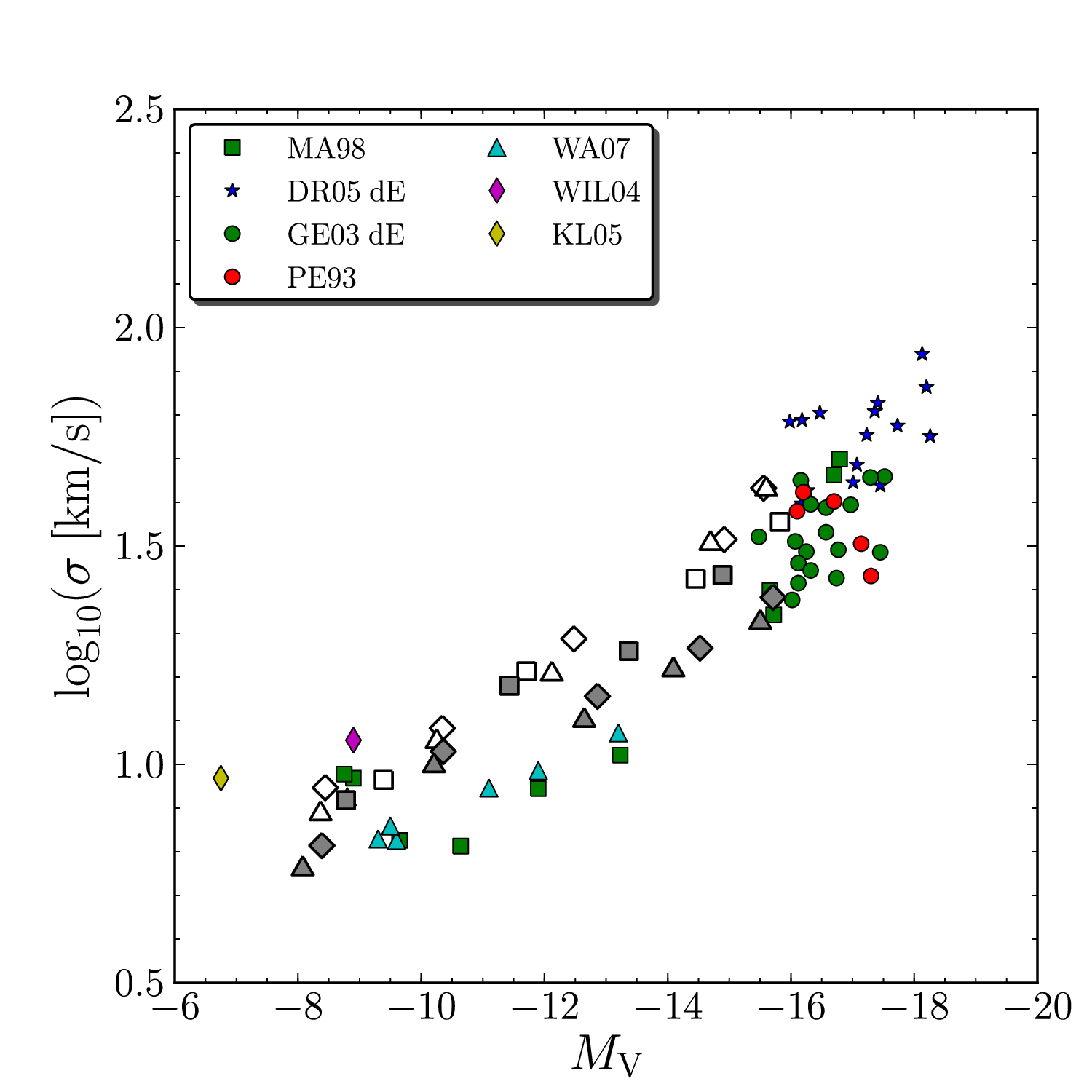}
 \caption{Stellar velocity dispersion versus V-magnitude. Observational
 data: \citealt{mateo98} (MA98), \citealt{rij05} (DR05), \citealt{ge03}
 (GE03), \citealt{peter93} (PE93), 7 MW dSphs from \citealt{walker07}
 (WA07), Ursa Minor from \citealt{wilkinson04} (WIL04), Ursa Major from
 \citealt{kleyna05} (KL05). \label{scaling_MV_disp}}
\end{minipage}
\end{figure*}

There is now a very strong difference in the correlation between the
local gas density and the separate stellar populations. The newest stars
are always found in the densest regions of the gas, which is not
unlogical considering the star formation criteria
\citep{sander:dgmodels}. The slightly older stars are much more likely
to be found in the bubbles or holes because individual groups of star
particles have had sufficient time to accumulate enough collective
feedback. This all speaks in favor of the multi-age model of
\citet{weisz09} for creating HI holes, and the findings of
\citet{stewart00} that young stars (H$\alpha$) prefer high density HI
regions while older stars (FUV) are more likely found in low density
regions. The idea that UV should be a better tracer for HI holes than
H$\alpha$ therefore seems very plausible. We can also spot cases of
triggered secondary star formation, the clearest example being at
$t=4.69$ where a large bubble at the lower left side expands outwards
and compresses the gas along a rim on the outside of the bubble,
spawning new star formation in this rim. Observational evidence for
similar events can be found where secondary SF is detected in H$\alpha$
along rims around HI holes \citep{stewart00, book08}.

All this again translates into the SFH characteristics we discussed
before for rotating dwarf galaxies, where the periodicity, or in other
words the large scale oscillation, of the SFHs from the spherical
models is significantly reduced. At times when in the spherical and
flattened models star formation has almost completely ceased, the
rotating model still shows a significant activity. It continuously
forms stars throughout the entire simulation. 

As a last point, the SF is also noticeably more spatially extended
than in the spherical and flattened cases. Moreover, the spatial
extent is quite constant during the entire simulation. Stars are
always formed throughout practically the entire body of the galaxy,
while in the spherical/flattened cases the subsequent SF bursts become
increasingly centrally concentrated.

\begin{figure*}
\begin{minipage}[t]{\columnwidth}
\centering
 \includegraphics[width=\columnwidth]{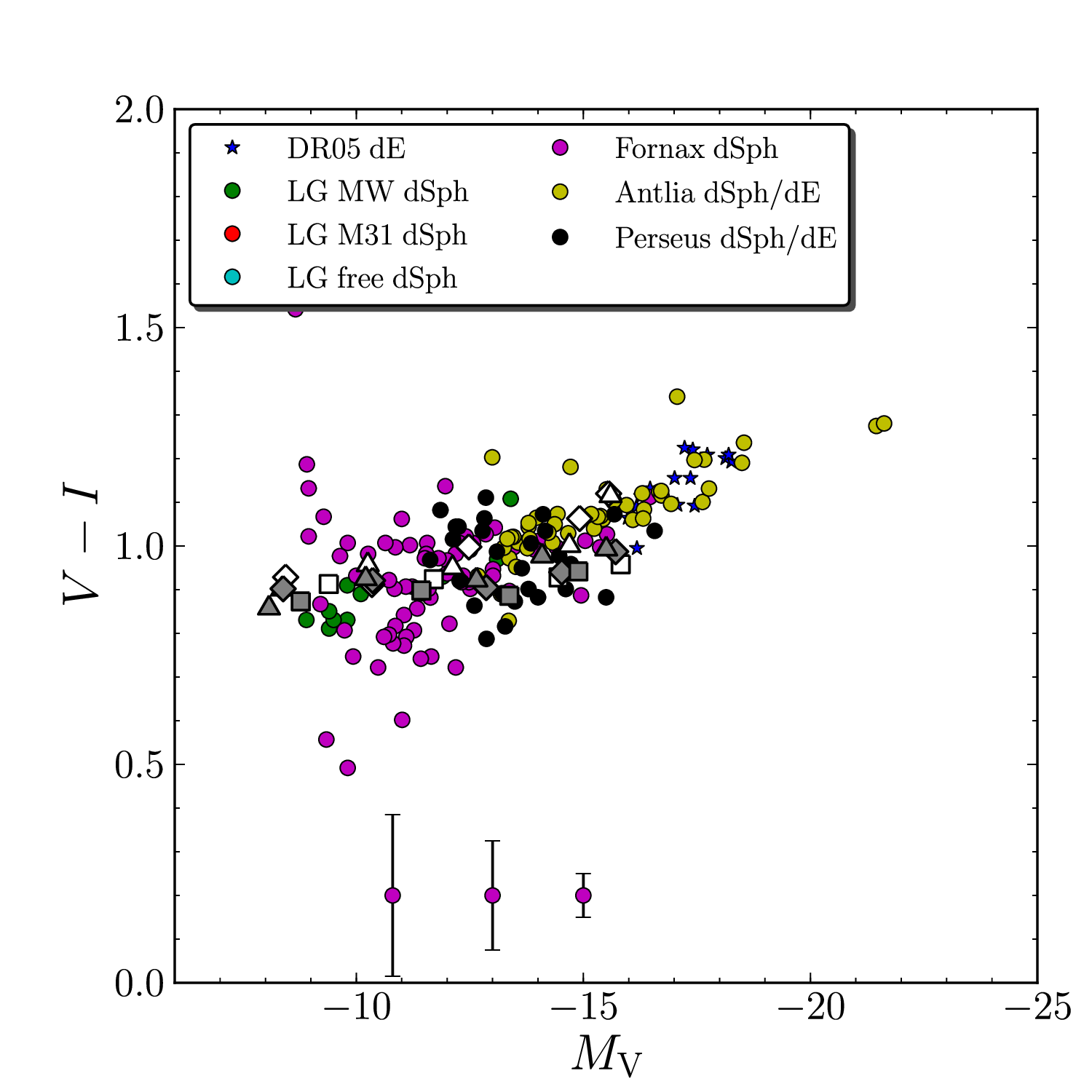}
 \caption{$V-I$ color versus V-magnitude. Symbols as in Fig.
 \ref{scaling_MV_re}, and the typical error bars for the Fornax Cluster
 dSph data are shown. Observational data as in
 Fig. \ref{scaling_MV_re}. \label{scaling_MV_color}}
\end{minipage}
\hfill
\begin{minipage}[t]{\columnwidth}
\centering
 \includegraphics[width=\columnwidth]{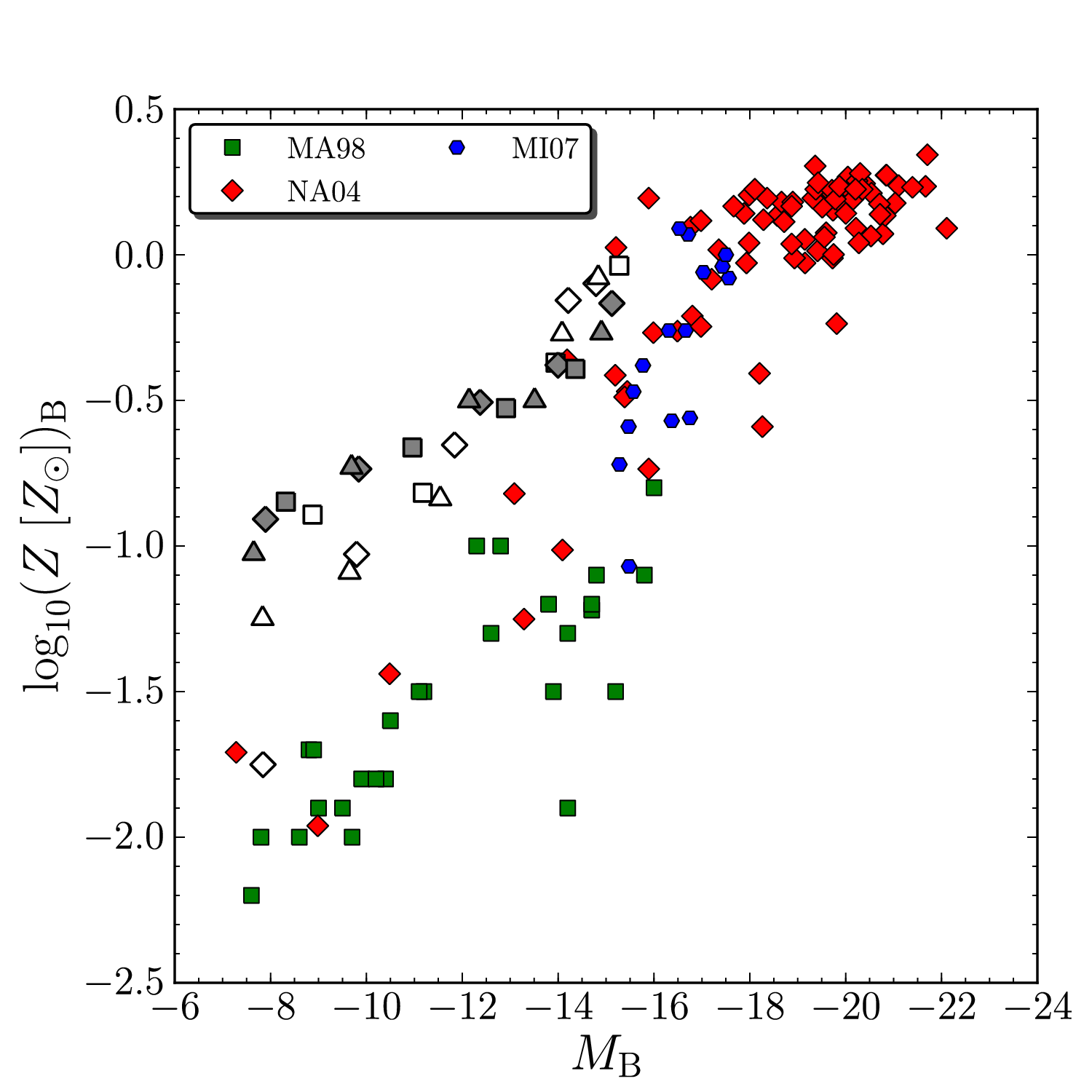}
 \caption{Metallicity in $\log_{10}(Z\ [Z_\odot])_\mathrm{B}$ (weighed
 with B-band luminosity, with $\mathrm{Z_\odot}=0.02$) versus
 B-magnitude. Data : \citealt{mateo98} (MA98), \citealt{naga04} (NA04),
 \citealt{mich07} (MI07).  \label{scaling_MB_Z}}
\end{minipage}
\end{figure*} 

\subsection{Scaling relations}
\label{section_scaling_relations}
 
Aside from the specific characteristics of individual models we
discussed above, we also consider the global photometric and
kinematical scaling relations traced by the simulated galaxies.  Our
main aim is to see how well the general characteristics of our
simulated dwarf galaxies agree with observational data of dwarf
galaxies as a class.

In the following, two series of simulations are plotted. Firstly, all
non-rotating galaxy models, both with spherically symmetric and with
flattened halos, represented with white symbols in
Figs. \ref{scaling_MV_re} to \ref{scaling_FP_LB}. Secondly, all galaxy
models initially rotating at $v_{rot} = 5$ km/s, both with spherically
symmetric and with flattened halos, are represented with grey symbols in
these figures. The symbol shapes distinguish the initial halo
flattenings: simulations with $q=1$ are shown with lozenges, $q=0.5$
with triangles, $q=0.1$ with squares. See Tables \ref{simulations} and
\ref{simulations_summary}.

\subsubsection{Half-light radius $R_{e}$}

The effects of flattening and rotation on the half-light radius are
clear in Fig. \ref{scaling_MV_re}, where $R_{e}$ versus $M_{V}$ is
shown.  Overall, at a fixed luminosity, rotation causes $R_{e}$ to
increase. The initial flattening of the halo does not seem to make a
significant difference since both model sequences are quite narrow. Only
at the high mass end of the non-rotating series does $R_{e}$ increase
with flattening, the rotating series are unaffected. Overall, there does
not seem to be a second parameter effect. Both series together nicely
encompass the observational width of the scaling relation. The third
series of simulations (with a low initial rotation speed of $v_{rot} =
1$ km/s) were omitted for clarity of the plot. They simply lie between
the two plotted series, providing us with rotation as a possible
explanation for the width of the scaling relation.

\subsubsection{Velocity dispersion $\sigma$}

Fig. \ref{scaling_MV_disp} shows the stellar central velocity dispersion
$\sigma$ versus $M_{V}$, projected along the line of sight. We take this
to be the $x$-axis, viewing the models edge-on. As in
\citet{sander:dgmodels}, the central velocity dispersion is in general
somewhat too high. However, the rotating models, having lower velocity
dispersions, compare favorably to spherically symmetric or flattened
ones. This decrease of the velocity dispersion in the more massive
models is tied to the increase of the half-light radius in the more
massive models. In the rotating series the initial flattening also
appears to have somewhat of an effect, leading to slightly higher
velocity dispersions in the most flattened cases (squares). This is
possibly due to the high M/L ratio of these systems, since they have a
considerably higher dynamical mass (for a given stellar mass) than their
initially less flattened counterparts (see Table
\ref{simulations_summary}).

\subsubsection{Color $V-I$}

The global $V-I$ colors of the models are shown in Fig.
\ref{scaling_MV_color}. The rotating galaxies lie a little lower than
the non-rotating on this plot, meaning these galaxies are slightly
bluer. This can be understood from their star formation histories
(Fig. \ref{SFH_rot}). The strength of the first SF peak is reduced,
while at later times SF is enhanced with respect to the non-rotating
case, producing more younger, bluer stars. Otherwise, all simulations
fall well within the observational range. This is, however, not a very
stringent test of the models, given the fact that the $V-I$ color of
an intermediate-age stellar population is relatively insensitive to
metallicity (see next paragraph).

\subsubsection{Metallicity}

\label{section_scaling_metallicity}

The metallicity of all dwarf models is too high, especially in the
low-mass regime. This problem was already encountered by
\citet{sander:dgmodels} for the spherically symmetric models. Below $M_B
\approx -12$~mag, the rotating models are more metalrich than
non-rotating ones whereas above this magnitude they are less metalrich.
An explanation for the low mass models can again be found in the
respective SFHs (Fig. \ref{SFH_rot}). In the least massive non-rotating
models, the large first peak in the SFH strongly inhibits further star
formation because the combined force of the feedback is strong enough to
severely lower the gas density. Adding rotation reduces this first peak
and thus also its truncating power, allowing SF to proceed continuously
and enrich the gas further with subsequent stellar generations. When
going to higher masses however, the effect and importance of the first
peak decreases. From the C07 model on, the trend reverses. This is most
likely due to the strong decrease in central concentration of SF and
feedback because of rotation (see Figs. \ref{snapshots} and
\ref{SFR_radius} at the end of this paper), together with the simple
fact that less stellar mass is produced. SF, metal production and gas
enrichment are much more diffuse, providing (on average) less-metalrich
gas for subsequent stellar generations.

\subsubsection{Surface brightness profiles}

\begin{figure}
\centering
 \includegraphics[width=\columnwidth]{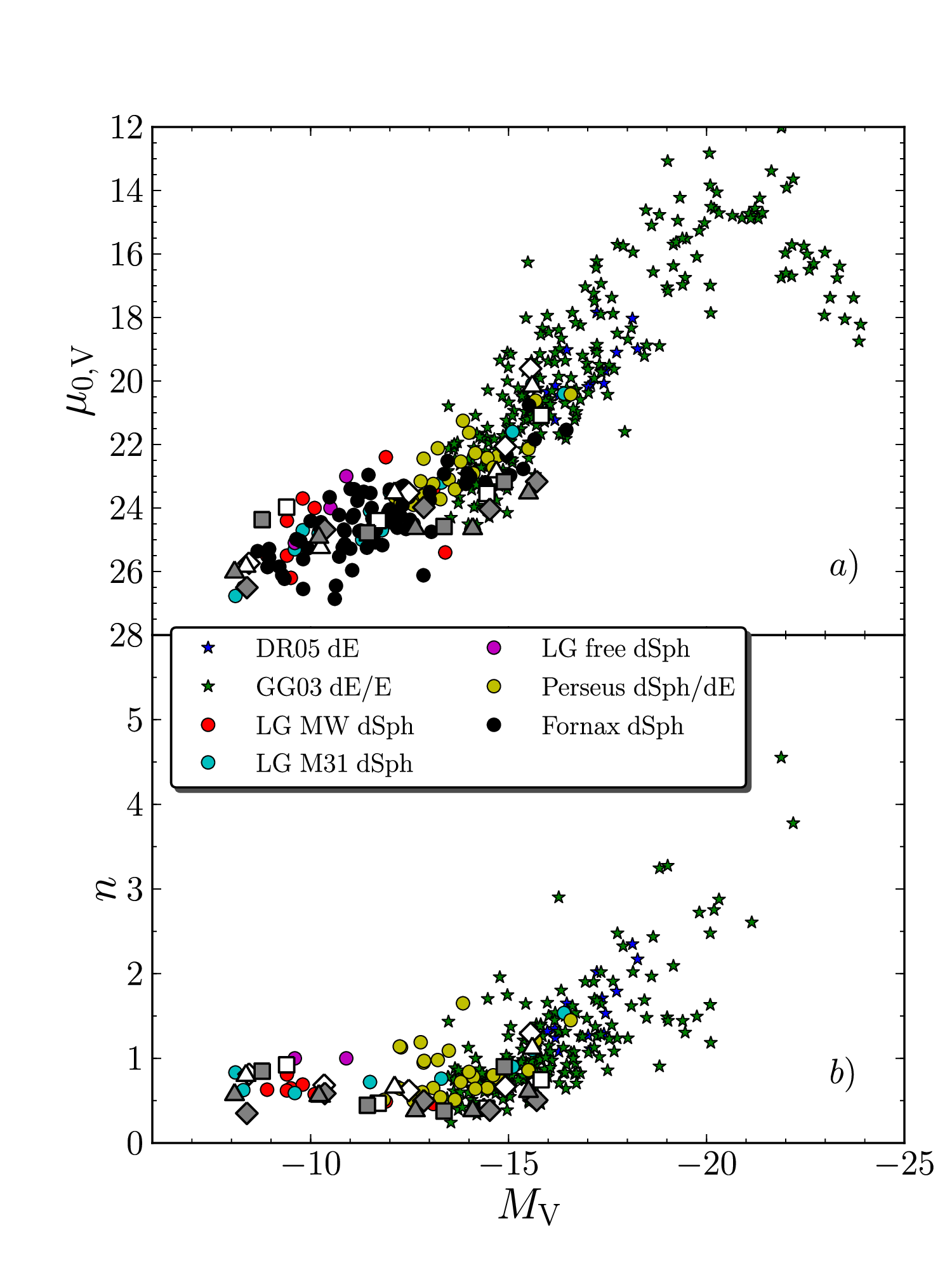}
 \caption{S\'ersic parameters versus V-magnitude. {\em Upper
 panel}: central surface brightness $\mu_{0}$ in the V-band, {\em lower
 panel}: S\'ersic index $n$. Observational data as in
 Fig. \ref{scaling_MV_re}.}

 \label{scaling_MV_sersic}

\end{figure}

The surface brightness profiles are fitted with a S\'ersic law 
\begin{equation}
 I(R)=I_{0}e^{-\left(\frac{R}{R_{0}}\right)^{1/n}},
\end{equation}
from which the parameters $\mu_{0}$ and $n$ are plotted and compared
with observational data in Fig. \ref{scaling_MV_sersic}. The
S\'ersic index $n$ does not differ significantly between rotating and
non-rotating models. The central surface brightness $\mu_{0}$, on the
other hand, is consistently lower in the rotating models. This is to
be expected:~with rotation the SF becomes less centrally concentrated
and more widespread, lowering the central surface brightness.

\subsubsection{Fundamental plane}

\begin{figure}
\centering
 \includegraphics[width=\columnwidth]{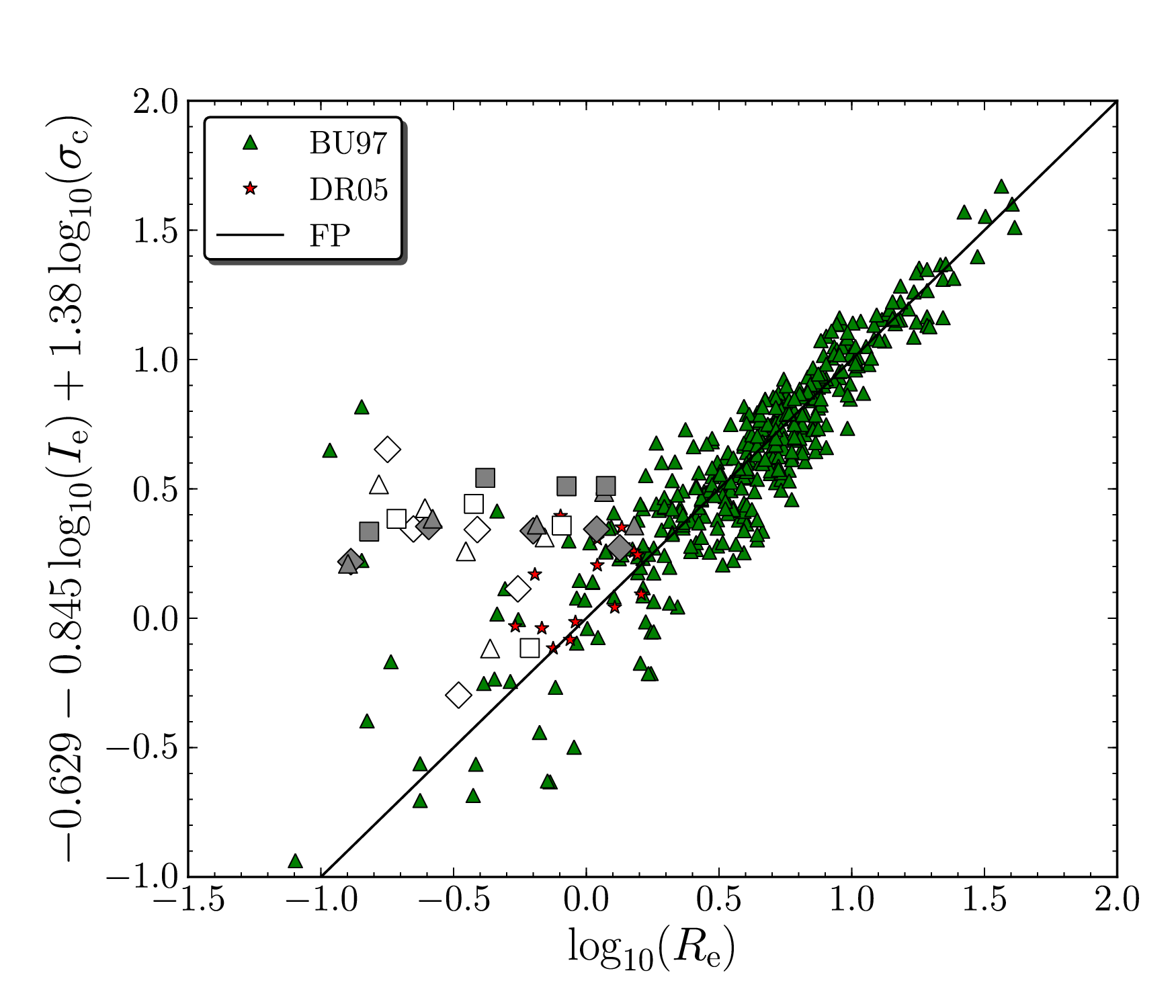}
 \caption{Side-view of the fundamental plane in physical
 coordinates. White symbols are our non-rotating galaxies; the grey ones
 are the fastest rotating galaxies, shapes denote initial flattening as
 in Fig. \ref{scaling_MV_re}. All other points on the plot are
 observational data, taken from \citealt{burstein97} (BU97) and
 \citealt{rij05} (DR05).}  \label{scaling_FP_Re}

\end{figure}

\begin{figure}
\centering
 \includegraphics[width=\columnwidth]{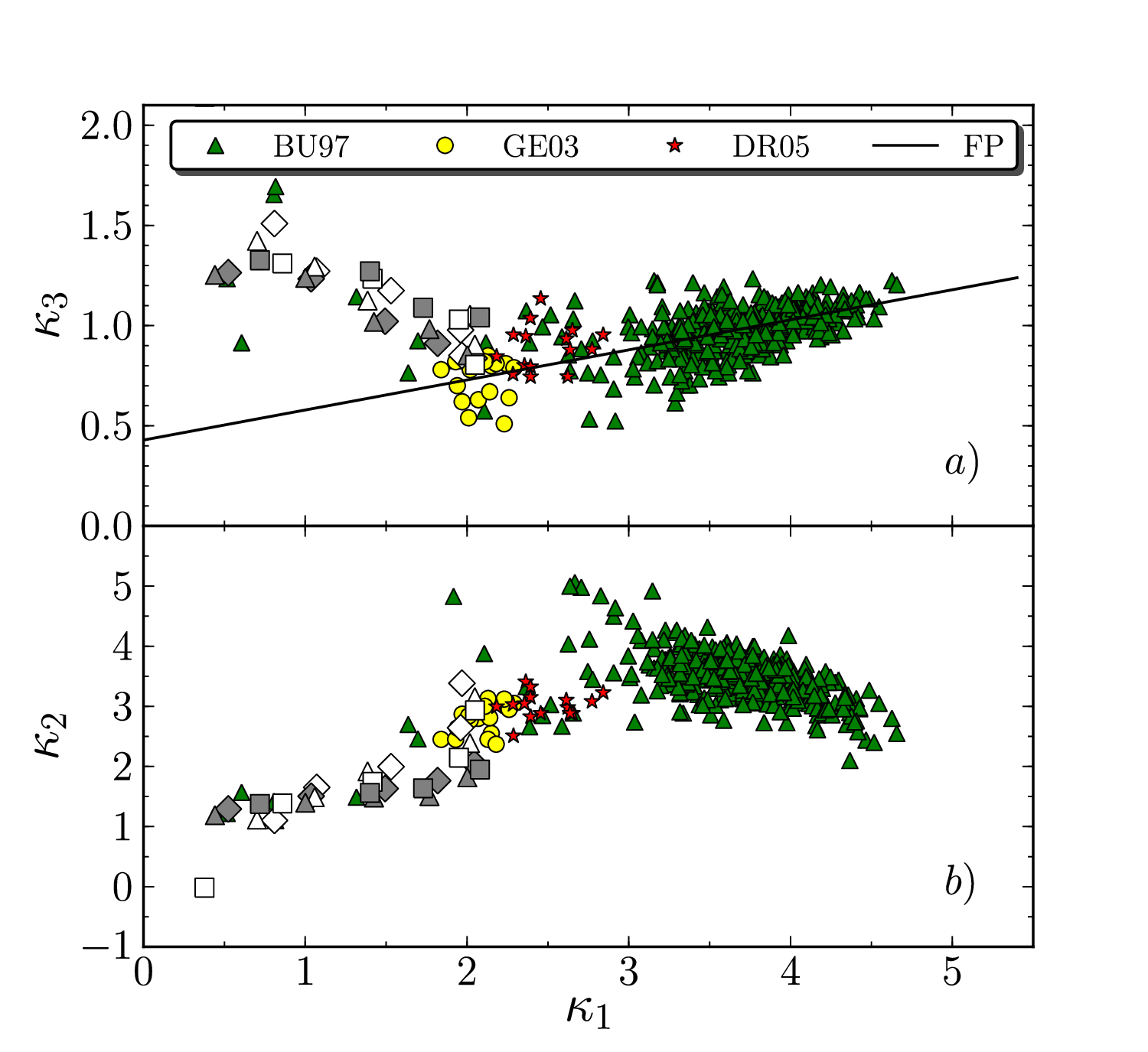}
 \caption{The fundamental plane in $\kappa$-space. {\em Upper panel}:
 side view of fundamental plane ($\kappa_{3}$, $\kappa_{1}$), {\em lower
 panel}: face-on view ($\kappa_{2}$, $\kappa_{1}$). Observational data
 from \citealt{ge03} (GE03), other symbols as in Fig.
 \ref{scaling_FP_Re}.}

 \label{scaling_FP_kappa}

\end{figure}

\begin{figure}
\centering
 \includegraphics[width=\columnwidth]{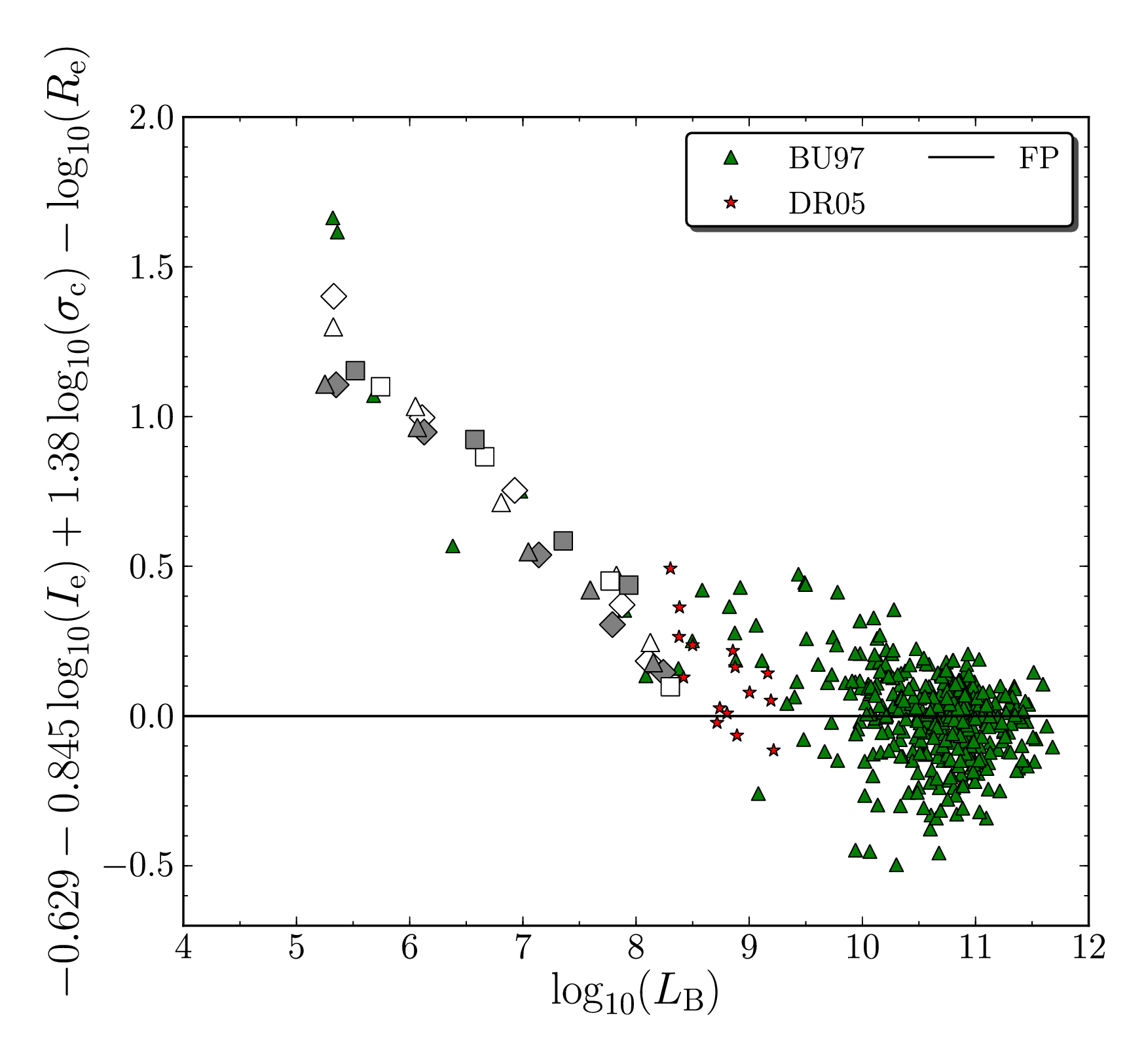}
 \caption{Deviation from the fundamental plane. Symbols as in Fig.
 \ref{scaling_FP_Re}.}

 \label{scaling_FP_LB}

\end{figure}

The fundamental plane \citep{bender1992, burstein97} is shown in
physical coordinates (Fig.  \ref{scaling_FP_Re}) and in $\kappa$ space
(Fig.  \ref{scaling_FP_kappa}). The ``vertical'' deviation from the
fundamental plane is shown in Fig.  \ref{scaling_FP_LB}. Except for the
most massive non-rotating galaxy models (white lozenges), which are very
compact, most dwarf galaxy models lie significantly above the
fundamental plane. These compact non-rotating dwarfs have small $R_e$
and consequently high mean surface brightness within $R_e$ (denoted by
$I_e$), making them stick out in the side-view of the fundamental plane
(Fig. \ref{scaling_FP_Re}) and in its $\kappa_1-\kappa_2$ projection
(Fig. \ref{scaling_FP_kappa}).

Overall, the simulations agree very well with the observational trends
and luminosity dependent deviations from the fundamental plane
(observational data taken from \citealt{burstein97}). Even the compact
non-rotating models fall within the observational spread of the
relation.

\section{Results / Discussion}
\label{section:results}

\subsection{Evaluation of analysis}

From the previous paragraph, it is clear that rotation has a more
pronounced influence on the observational properties of the simulated
dwarf galaxies, quantified by photometric and kinematical scaling
relations, than the flattening of the initial conditions. The
differences {\em between} the sequences of rotating and non-rotating
models are significantly larger than those between flattened and
non-flattened galaxies {\em within} each sequence. Still, all models
fall within the range allowed by the data, apart from the problems we
noted with metallicities being to high. Moreover, despite their
simplicity, this suite of simulations suggests a possible explanation
for the widths of the observed scaling relations. While mass is the
dominant parameter that determines the shape and slope of each scaling
relation, angular momentum could be an important second parameter that
determines the width of the relations. This will however not be the only
factor, since external influences such as environment and merger history
are likely to also have a significant influence here. And we should also
note the inherent variance that is present in our models, as discussed
in section \ref{variance_eval} and shown in Fig. \ref{variance}.

While the effects of flattening and rotation on the observed scaling
relations are modest, the addition of rotation has a strong effect on
the details of the evolution of dwarf galaxies. This is most clearly
seen in the properties of the stellar populations, e.g. in the
metallicity profiles (Figs. \ref{metprof_rot} and \ref{metprof_flat}),
SFHs (Figs. \ref{SFH_rot} and \ref{SFH_flat}), and overall appearances
(Fig. \ref{snapshots}). In this respect, rotating models are
qualitatively quite distinct from non-rotating ones, independent of
initial flattening:~rotating models have continuous SFHs with
wide-spread SF while non-rotating models have ``breathing'' SFHs with
centrally concentrated SF. Observationally, this leads to rotating
models having flat metallicity profiles while non-rotating models show
pronounced negative metallicity gradients.

\subsection{Mechanism}
\label{section_mechanism}

Within the sequence of non-rotating models, flatttening of the initial
conditions appears to have very little effect on the models' properties,
which is of importance especially when considering the metallicity
profiles. It therefore seems doubtful that the fountain mechanism is
very important for dwarfs. Still, if there are large feedback driven
outbursts of gas, they tend to be aligned preferentially along the minor
axis. But the expelled enriched gas does not fall back onto the
galaxy. This is most likely because of the shallow potential wells of
dwarf galaxies, and because the remaining cold gas is simply ``in the
way''. Another important argument against the fountain mechanism is the
actual flattening of dwarf galaxies, both in observations and in our
simulations, as we discussed in section \ref{flattening_eval}.  Dwarf
galaxies simply are not likely to occur with very flat shapes
\citep{sanchez10}. Their flattenings are not comparable to those of
massive spiral galaxies ($q\approx0.2$), they are on average much
thicker ($\langle q \rangle \approx 0.6$). This often makes it difficult
to even speak of a ``disk'' in the context of dwarf galaxies. Therefore,
while the fountain mechanism might be very relevant in the domain of
large spiral galaxies, with much deeper potential wells and much flatter
shapes, it does not appear an important mechanism in dwarf galaxies.

\begin{figure*}
\begin{minipage}{180mm}

 \centering{}
 \includegraphics[width=0.8\textwidth]{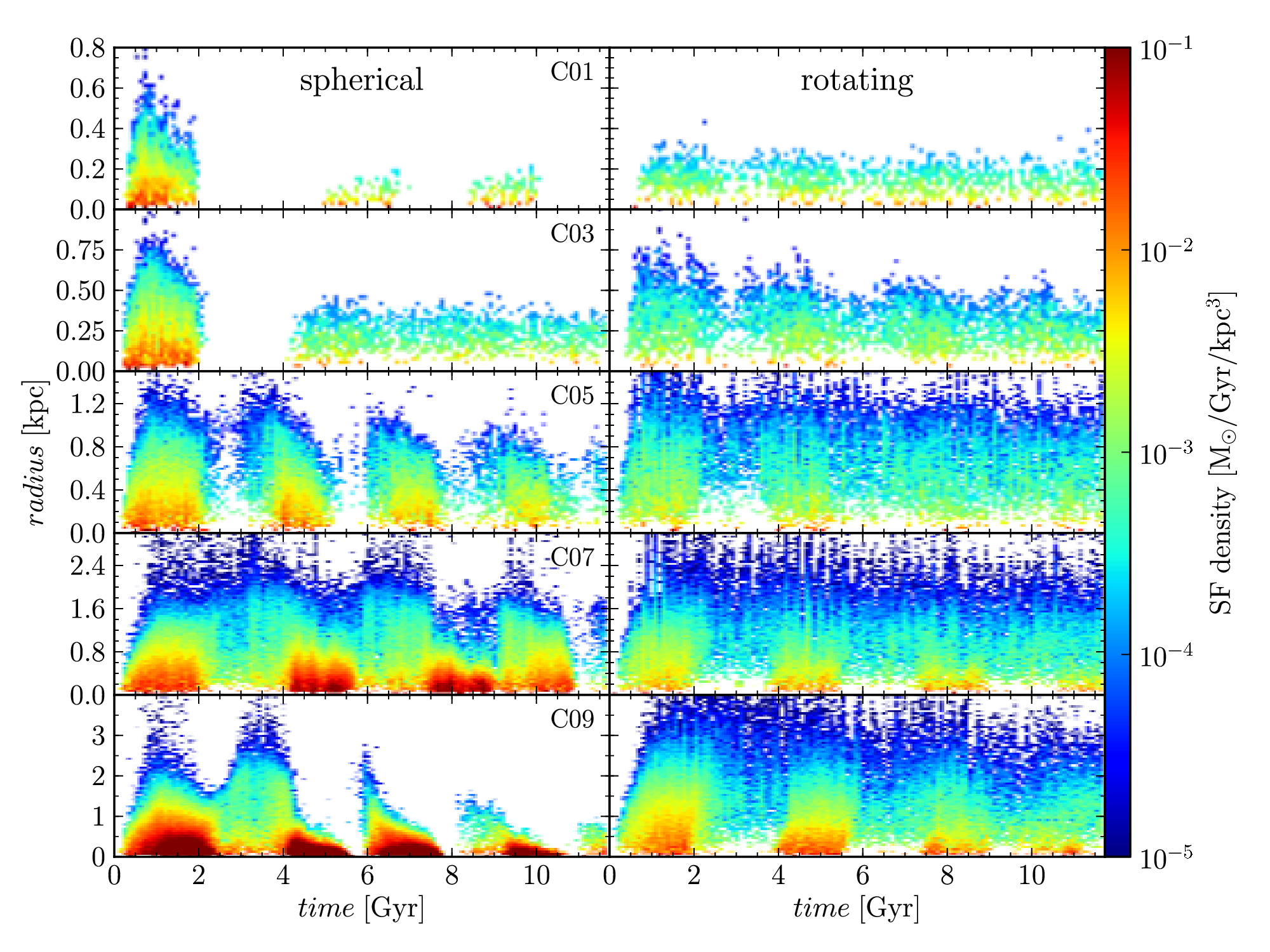}
 \caption{Star formation density in $\mathrm{M_\odot / Gyr / kpc^{3}}$
 for spherical, non-rotating ({\em left column}) and rotating models
 ({\em right column}). The total mass of the models increases from the
 top down as indicated in the figure. SF density is plotted in
 color-code versus time on the $x$ axis and radius on the $y$ axis.}

 \label{SFR_radius}

\end{minipage}
\end{figure*}

Rotation, on the other hand, leads to important qualitative and
quantitative changes in the SFHs of dwarfs. The consequences of the
addition of angular momentum are the following:

\begin{enumerate}
 \item Gas will spiral inward, instead of falling straight to the
       center. There is a ``{\em centrifugal barrier}''
   preventing the gas from collapsing to a dense central region.
 \item Since the gas density is much more smeared out, so is the star
   formation. The density criterion (see section \ref{section_SF}) for
   star formation is now reached in a much larger region of the gas,
   so that star formation will occur throughout practically the entire
   body of the galaxy. This is evident in Figs. \ref{snapshots} and
   \ref{SFR_radius}:~star formation is consistently more spatially
   extended in comparison with non-rotating models.
 \item This naturally produces more spatially homogeneous stellar
   populations. Therefore the gas is enriched much more homogeneously
   across the entire galaxy, explaining the flat metallicity profiles
   in Fig. \ref{metprof_rot}.
 \item Where there is star formation, unavoidingly there will also
   be stellar feedback. Since the former is smeared out across almost
   the entire galaxy, so is the latter. The supernova feedback now being
   less centrally concentrated, this leads to much less pronounced
   large-scale collective behaviour of the gas. The effects of feedback
   are now more {\em local}. This has two distinct but related effects:
   \begin{itemize}
    \item The supernovae combine their energy locally on a smaller
	  scale, and produce {\em low-density holes} in the gas,
	  instead of collectively blowing out the gas and lowering the
	  global gas density after a large centralized star formation
	  event. This hole or bubble structure is clearly visible in
	  Fig. \ref{snapshots} and is discussed in section
	  \ref{section_gas_structure}.
    \item This can also be linked to our findings concerning the
          SFHs in section \ref{section_SFH} and Fig. \ref{SFH_rot}. Since the
          gas does not collectively blow out due to feedback, star
          formation will not shut down completely across the entire
          galaxy, because only locally the density criterion for star
          formation is not satisfied (in the feedback holes,
          Fig. \ref{snapshots}). This is also seen in
          Fig. \ref{SFR_radius}. Collective behaviour -
          i.e. large-scale oscillations in the SFR - is diminished,
          leading to more continuous, less variable SFH. The
          ``breathing'' SF, typical of non-rotating models, is largely
          absent.
\end{itemize}
\end{enumerate}

The density criterion mentioned in section \ref{section:code} and here
in point (ii) is an important element of our models. We should note that
we employ a treshold of $0.1 ~\mathrm{cm^{-3}}$, while
\citet{governato10} suggest the usage of a treshold of $100 ~%
\mathrm{cm^{-3}}$, reflecting more realistically the conditions of real
star-forming gas clumps. We do not expect this to qualitatively change
the proposed mechanism however: the higher treshold will take the gas
longer to reach it when collapsing, but the extension of the cooling
curves below $10^{4}$K \citep{maio07} will cause the gas to collapse
easier and on smaller scales. These effects might not cancel each other
out, but the relative influence of added rotation, as discussed above,
will remain qualitatively similar. Perhaps on small scales the chemical
homogeneity will be less, but on large scales rotating galaxies will
still be chemically homogeneous. The inclusion of the high-density
treshold and the extra cooling will be the subject of further research.

\subsection{Galaxy types}

As already mentioned in paragraph \ref{section_scaling_relations}, our
dwarf galaxy models agree quite well with the observed the scaling
relations of early-type galaxies. However, since the model galaxies
still contain gas and have ongoing star formation at the end of the
simulation, they should be classified as late-type dwarfs.

\subsubsection{dIrrs?}

The non-rotating and slowly rotating models, both flattened and
non-flattened, are characterized by
\begin{itemize}
\item centrally concentrated gas distribution; high central density
\item low specific angular momentum
\item strong stellar population gradients
\item bursty or episodic SF
\item centrally concentrated SF
\item large-scale feedback driven outflows and a largely featureless
  ISM.
\end{itemize}
The fast rotating models, both flattened and non-flattened, are
characterized by
\begin{itemize}
\item spatially extended gas distribution; low central density
\item high specific angular momentum
\item small stellar population gradients, if any
\item continuous SF
\item small star forming regions, scattered across the galaxy
\item turbulent ISM with distinct feedback driven holes.
\end{itemize}

dIrrs are known to have a more extensive and less centrally concentrated
gas distribution that other gas-rich dwarf galaxy types (e.g BCDs), and
also a relatively high specific angular momentum. Chemical homogeneity
is a general trait of dIrrs, both in their gas and their stellar content
\citep{tolstoy:dgreview,kobulnicky-skillman:dirrmetprof}. From the
review of dwarf galaxy properties in \citet{tolstoy:dgreview} and the
extensive work of \citet{dolphin:dgsfh}, using CMD analysis to
reconstruct dwarf galaxy SFHs
(\citealt{aparicio:cmdanal,dolphin:cmdanal1,dolphin:cmdanal2,tolstoy:cmdanal,tosi:cmdanal}),
it is clear that dIrrs generally have a ``continuous'' SFH without
quiescent periods without SF. The characteristic gas structures of dIrrs
have already been discussed in section \ref{section_gas_structure}.

From this short overview of the observed properties of late-type dwarfs,
it is clear that our fast rotating models resemble dIrrs, at least
qualitatively. But our non-rotating and slowly rotating models do not,
although they do also still contain gas and show ongoing (periodic) star
formation. Angular momentum, it seems, invokes different star formation
modes in dwarf galaxies. It differentiates between centralized/bursty
and extended/continuous star formation, and all dwarf galaxy properties
connected with this which are mentioned above. Although quantitatively
not comparable to our models, it is worth mentioning BCDs and their
differences with dIrrs. They too are gas-rich late-type dwarf galaxies,
but have a lower specific angular momentum and much more concentrated
gas distribution \citep{vanzee01,vanzee02}. BCDs also show substantial
color gradients \citep{vanzee02}, indicating chemical inhomogeneity, and
by definition have bursting SFHs.

\subsubsection{Conversion of late-type dwarfs to early types}

Since internal processes such as supernova feedback are not capable of
removing the gas from a dwarf galaxy, we turn to external or
environmental processes, e.g. tidal stripping and ram pressure
stripping \citep{ma06}. Ram-pressure stripping is able to remove a
large fraction of the gas and leaves the structure and kinematics of
the stars relatively undisturbed, thus preserving any pre-existing
stellar population gradients and rotation \citep{gre03,marcolini2003}.
Tidal interactions can cause violent reactions in dynamically cold
thin-disk dwarf galaxies and can significantly disturb them
\citep{mayer01a,mayer01b}. However, the majority of the dwarf
late-type population is quite round, with mean axis ratio $\langle q
\rangle \approx 0.6$. In such galaxies, tidal interactions wreak much
less havoc \citep{sander:phd}.

We therefore argue that it is possible to convert late-type dwarfs
into early-type ones inside a cluster environment by removing their
gas and halting SF without significantly altering their structural and
kinematical properties. So the rotation which is present in dIrrs can
be preserved in their dE descendants along with the stellar
characteristics connected with rotation (metallicity profiles).

\section{Conclusion}
\label{section:conclusion}

The centrifugal barrier mechanism formulated in section
\ref{section:results}, is able to combine all our findings we discussed
in the analysis into one coherent picture, emphasizing the
importance of rotation in dwarf galaxy behaviour.

\subsection{Metallicity profiles}

Our interest in this subject was triggered initially by the finds of
\citet{koleva-sven:demetprof}, who found that dwarf early-type galaxies
without stellar population gradients were also the fastest rotating
ones. We conclude from our simulations that (in isolation) rotation, or
the absence thereof, is indeed a key factor in creating stellar
population gradients. The ``fountain mechanism'' does not seem relevant
on the scale of dwarf galaxies, and our simulations clearly indicate
that the {\em geometry} or {\em flattening} of a dwarf galaxy does not
have any significant influence:~pressure-supported, non-rotating systems
behave very much alike, independent of flattening.

We therefore propose the alternative ``centrifugal barrier mechanism''
in section \ref{section_mechanism}, which explains the existence of
flat metallicity profiles as a natural consequence of its rotation.

\subsection{Angular momentum as second parameter}

We suggest angular momentum as being a crucial {\em second parameter}
in determining the appearance and evolution of dwarf galaxies, with
the total galaxy mass being the prominent first parameter. While our
simulations are admittedly very idealized and cannot purport to paint
a cosmologically up-to-date picture of dwarf galaxy formation, they
have the enormous benefit of allowing us to unambiguously identify the
influence of individual parameters, such as angular momentum.

We have shown that rotation has a significant impact on the stellar
populations of dwarf galaxies. And in the same vein we can say the
opposite for dwarf galaxy flattening, which shows no significant
influence in our simulations, and thus is less likely to be a major
player in dwarf galaxy evolution.

\subsection{Making dIrrs}

We find that without rotation, it does not seem possible to
qualitatively produce ``typical'' dIrrs with spatially extended SF,
continuous SFHs, a turbulent ISM with low-density holes, and most
importantly with chemical homogeneity throughout its body of gas and
stars. Non-rotating models do not display any of these characteristics
(having centralized SF, bursty SFHs, featureless ISMs and metallicity
gradients). Angular momentum appears to differentiate between bursty and
continuous star formation modes.

\section*{Acknowledgements}
We wish to thank the anonymous referee for the many helpful comments and
suggestions which improved this paper. We also thank Volker Springel for
making publicly available the {\sc Gadget-2} simulation code. JS would
like to thank the Fund for Scientific Research - Flanders (FWO). All
simulations were run on our local computer cluster ITHILDIN.



\label{lastpage}


\begin{thebibliography}{}


\bibitem[\protect\citeauthoryear{Alard}{2001}]{alard01} Alard C., 2001,
		A\&A, 377, 389

\bibitem[\protect\citeauthoryear{Aparicio et
		al.}{1996}]{aparicio:cmdanal} Aparicio A., Gallart C.,
		Chiosi C., Bertelli G., 1996, ApJ, 469, L97

\bibitem[\protect\citeauthoryear{Barazza et al.}{2002}]{ba02} Barazza
  F. D., Binggeli B., Jerjen H., 2002, A\&A, 391, 823

\bibitem[\protect\citeauthoryear{Barazza \& Binggeli}{2002}]{binggeli}
			Barazza F.~D., Binggeli B., 2002, A\&A, 394, L15

\bibitem[\protect\citeauthoryear{Battaglia et al.}{2006}]{battaglia06}
		Battaglia G., et al., 2006, A\&A, 459, 423

\bibitem[\protect\citeauthoryear{Battaglia et al.}{2010}]{battaglia10}
		Battaglia G., Tolstoy E., Helmi A., Irwin M., Parisi P.,
		Hill V., Jablonka P., 2010, MNRAS, 1817

\bibitem[\protect\citeauthoryear{Bender, Burstein, \&
			Faber}{1992}]{bender1992} Bender R., Burstein
			D., Faber S.~M., 1992, ApJ, 399, 462

\bibitem[\protect\citeauthoryear{Bernard et al.}{2007}]{bernard:ic1613}
			Bernard E.~J., Aparicio A., Gallart C.,
			Padilla-Torres C.~P., Panniello M., 2007, AJ,
			134, 1124

\bibitem[\protect\citeauthoryear{Binggeli et al.}{1987}]{bi87} Binggeli
			B., Tammann G. A., Sandage A., 1987,
		AJ, 94, 251

\bibitem[\protect\citeauthoryear{Binggeli \&
			Popescu}{1995}]{binggeli:flatdist} Binggeli B.,
			Popescu C.~C., 1995, A\&A, 298, 63

\bibitem[\protect\citeauthoryear{Book, Chu, \& Gruendl}{2008}]{book08}
			Book L.~G., Chu Y.-H., Gruendl R.~A., 2008,
			ApJS, 175, 165

\bibitem[\protect\citeauthoryear{Burstein et al.}{1997}]{burstein97}
			Burstein D., Bender R., Faber S., Nolthenius R.,
			1997, AJ, 114, 1365

\bibitem[\protect\citeauthoryear{Buyle et al.}{2005}]{buy05} Buyle P.,
		De Rijcke S., Michielsen D., Baes M., Dejonghe H., 2005, MNRAS, 360,
		853

\bibitem[\protect\citeauthoryear{Caldwell}{1999}]{caldwell1999} Caldwell
			N., 1999, AJ, 118, 1230

\bibitem[\protect\citeauthoryear{Cioni}{2009}]{cioni:smc} Cioni
			M.-R.~L., 2009, A\&A, 506, 1137

\bibitem[\protect\citeauthoryear{Conselice et al.}{2003}]{co03}
  Conselice C. J., O'Neil K., Gallagher J .S. {\sc iii}, Wyse
  Rosemary F. G., 2003, ApJ, 591, 167

\bibitem[\protect\citeauthoryear{C\^ot\'e et al.}{2000}]{co00}
  C\^ot\'e S., Carignan C., Freeman K. C., 2000, AJ, 120, 3027

\bibitem[\protect\citeauthoryear{C\^ot\'e et al.}{2009}]{co09} C\^ot\'e
		S., Draginda A., Skillman E. D., Miller B. W., 2009, AJ,
		138, 1037

\bibitem[\protect\citeauthoryear{Dejonghe \& de Zeeuw}{1988}]{kuzmin}
			Dejonghe H., de Zeeuw T., 1988, ApJ, 333, 90

\bibitem[\protect\citeauthoryear{De Looze et al.}{2010}]{lo10} De
  Looze I., Baes M., Zibetti S., Fritz J., Cortese L., Davies J. I.,
  Verstappen J., Bendo G. J., Bianchi S., Clemens M., et al., 2010,
  A\&A, 518, L54

\bibitem[\protect\citeauthoryear{De Rijcke et al.}{2003a}]{rij03a} De
			Rijcke S., Zeilinger W. W., Dejonghe H., Hau
			G. K. T., 2003, MNRAS, 339, 225

\bibitem[\protect\citeauthoryear{De Rijcke et al.}{2003b}]{rij03b} De
  Rijcke S., Dejonghe H., Zeilinger W. W., Hau G. K. T., 2003, A\&A, 400, 119

\bibitem[\protect\citeauthoryear{De Rijcke et al.}{2005}]{rij05} De
			Rijcke S., Michielsen D., Dejonghe H., Zeilinger
			W.~W., Hau G.~K.~T., 2005, A\&A, 438, 491

\bibitem[\protect\citeauthoryear{De Rijcke et al.}{2009}]{rij09} De
			Rijcke S., Penny S.~J., Conselice C.~J., Valcke
			S., Held E.~V., 2009, MNRAS, 393, 798

\bibitem[\protect\citeauthoryear{De Young \&
		Gallagher}{1990}]{deyoung90} De Young D.~S., Gallagher
		J.~S., III, 1990, ApJ, 356, L15

\bibitem[\protect\citeauthoryear{De Young \&
			Heckman}{1994}]{deyoung:fountain} De Young
			D.~S., Heckman T.~M., 1994, ApJ, 431, 598

\bibitem[\protect\citeauthoryear{den Brok et al.}{2011}]{brok2011} den
		Brok M., et al., 2011, arXiv, arXiv:1103.1218

\bibitem[\protect\citeauthoryear{Dolphin et al.}{2005}]{dolphin:dgsfh}
			Dolphin A.~E., Weisz D.~R., Skillman E.~D.,
			Holtzman J.~A., 2005, astro, arXiv:astro-ph/0506430

\bibitem[\protect\citeauthoryear{Dolphin}{1997}]{dolphin:cmdanal1}
		  Dolphin A., 1997, NewA, 2, 397

\bibitem[\protect\citeauthoryear{Dolphin}{2002}]{dolphin:cmdanal2}
		  Dolphin A.~E., 2002, MNRAS, 332, 91

\bibitem[\protect\citeauthoryear{Dufour \&
			Harlow}{1977}]{dufour-harlow:smc} Dufour R.~J.,
			Harlow W.~V., 1977, ApJ, 216, 706

\bibitem[\protect\citeauthoryear{Ferguson \& Binggeli}{1994}]{ferbin}
			Ferguson H. C. \& Binggeli B., A\&ARv, 1994, 6,
			67

\bibitem[\protect\citeauthoryear{Ferrara \& Tolstoy}{2000}]{tolstoy:bo}
			Ferrara A., Tolstoy E., 2000, MNRAS, 313, 291

\bibitem[\protect\citeauthoryear{Geha et al.}{2003}]{ge03} Geha M.,
  Guhathakurta P., van der Marel R. P., 2003, AJ, 126, 1794

\bibitem[\protect\citeauthoryear{Governato et al.}{2010}]{governato10}
			Governato F., et al., 2010, Natur, 463, 203

\bibitem[\protect\citeauthoryear{Graham et al.}{2003}]{gra03} Graham
			A. W., Jerjen H., Guzm\'an R., 2003, AJ, 126,
			1787

\bibitem[\protect\citeauthoryear{Grebel et al.}{2003}]{gre03} Grebel
			E. K., Gallagher J. S. {\sc iii}, Harbeck D.,
			2003, AJ, 125, 1926

\bibitem[\protect\citeauthoryear{Harbeck et al.}{2001}]{harbeck01}
			Harbeck D., et al., 2001, AJ, 122, 3092

\bibitem[\protect\citeauthoryear{Hern{\'a}ndez-Mart{\'{\i}}nez et
			al.}{2009}]{hernandez:ngc6822}
			Hern{\'a}ndez-Mart{\'{\i}}nez L., Pe{\~n}a M.,
			Carigi L., Garc{\'{\i}}a-Rojas J., 2009, A\&A,
			505, 1027

\bibitem[\protect\citeauthoryear{Hunter \&
			Elmegreen}{2006}]{hunter:flatdist} Hunter D.~A.,
			Elmegreen B.~G., 2006, ApJS, 162, 49

\bibitem[\protect\citeauthoryear{Irwin \&
			Hatzidimitriou}{1995}]{irwin1995} Irwin M.,
			Hatzidimitriou D., 1995, MNRAS, 277, 1354

\bibitem[\protect\citeauthoryear{Jerjen et al.}{2000}]{je00} Jerjen H.,
			Kalnajs A., Binggeli B., 2000, A\&A, 358, 845

\bibitem[\protect\citeauthoryear{Kaufer et al.}{2004}]{kaufer:sextansA}
			Kaufer A., Venn K.~A., Tolstoy E., Pinte C.,
			Kudritzki R.-P., 2004, AJ, 127, 2723 

\bibitem[\protect\citeauthoryear{Kaufmann, Wheeler, \&
		Bullock}{2007}]{kaufmann07} Kaufmann T., Wheeler C.,
		Bullock J.~S., 200, MNRAS, 382, 1187

\bibitem[\protect\citeauthoryear{Kim et al.}{2005}]{smcHI} Kim S.,
			Staveley-Smith L., Dopita M.~A., Sault R.~J.,
			Freeman K.~C., Lee Y., Chu Y.~-., 2005, astro,
			arXiv:astro-ph/0506224

\bibitem[\protect\citeauthoryear{Kim et al.}{1999}]{kim99} Kim S.,
			Dopita M.~A., Staveley-Smith L., Bessell M.~S.,
			1999, AJ, 118, 2797

\bibitem[\protect\citeauthoryear{Kleyna et al.}{2005}]{kleyna05} Kleyna
		J.~T., Wilkinson M.~I., Evans N.~W., Gilmore G., 2005,
		ApJ, 630, L141

\bibitem[\protect\citeauthoryear{Kobulnicky \&
			Skillman}{1997}]{kobulnicky-skillman:dirrmetprof}
			Kobulnicky H.~A., Skillman E.~D., 1997, ApJ,
			489, 636

\bibitem[\protect\citeauthoryear{Koleva et
			al.}{2009}]{koleva-sven:demetprof} Koleva M., de
			Rijcke S., Prugniel P., Zeilinger W.~W.,
			Michielsen D., 2009, MNRAS, 396, 2133

\bibitem[\protect\citeauthoryear{Lianou, Grebel, \&
		Koch}{2010}]{lianou10} Lianou S., Grebel E.~K., Koch A.,
		2010, A\&A, 521, A43

\bibitem[\protect\citeauthoryear{Lisker et al.}{2006}]{li06} Lisker T.,
			Glatt K., Westera P., Grebel E. K., 2006, AJ,
		132, 2432

\bibitem[\protect\citeauthoryear{Mac Low \& Ferrara}{1999}]{maclow:bo}
			Mac Low M.-M., Ferrara A., 1999, ApJ, 513, 142

\bibitem[\protect\citeauthoryear{Maio et al.}{2007}]{maio07} Maio U.,
			Dolag K., Ciardi B., Tornatore L., 2007, MNRAS,
			379, 963

\bibitem[\protect\citeauthoryear{Marcolini, Brighenti, \&
			D'Ercole}{2003}]{marcolini2003} Marcolini A.,
			Brighenti F., D'Ercole A., 2003, MNRAS, 345,
			1329

\bibitem[\protect\citeauthoryear{Masegosa, Moles \& del
			Olmo}{1991}]{masegosa:dirrs} Masegosa J., Moles
			M., del Olmo A., 1991, A\&A, 249, 505

\bibitem[\protect\citeauthoryear{Mateo}{1998}]{mateo98} Mateo M.~L.,
			1998, ARA\&A, 36, 435

\bibitem[\protect\citeauthoryear{Mayer et al.}{2001a}]{mayer01a} Mayer
			L., Governato F., Colpi M., Moore B., Quinn T.,
			Wadsley J., Stadel J., Lake G., 2001, ApJ, 559,
			754

\bibitem[\protect\citeauthoryear{Mayer et al.}{2001b}]{mayer01b} Mayer
			L., Governato F., Colpi M., Moore B., Quinn T.,
			Wadsley J., Stadel J., Lake G., 2001, ApJ, 547,
			L123

\bibitem[\protect\citeauthoryear{Mayer et al.}{2006}]{ma06} Mayer L.,
  Mastropietro C., Wadsley J., Stadel J., Moore B., 2006, MNRAS, 369,
  1021

\bibitem[\protect\citeauthoryear{McConnachie \& Irwin}{2006}]{mccon2006}
			McConnachie A.~W., Irwin M.~J., 2006, MNRAS,
			365, 1263

\bibitem[\protect\citeauthoryear{McConnachie, Arimoto, \&
			Irwin}{2007}]{mccon2007} McConnachie A.~W.,
			Arimoto N., Irwin M., 2007, MNRAS, 379, 379

\bibitem[\protect\citeauthoryear{McCray \& Kafatos}{1987}]{mccray87}
			McCray R., Kafatos M., 1987, ApJ, 317, 190

\bibitem[\protect\citeauthoryear{Michielsen et al.}{2007}]{mich07}
			Michielsen D., et al., 2007, ApJ, 670, L101

\bibitem[\protect\citeauthoryear{Mieske et al.}{2007}]{mieske2007}
			Mieske S., Hilker M., Infante L., Mendes de
			Oliveira C., 2007, A\&A, 463, 503

\bibitem[\protect\citeauthoryear{Nagashima \& Yoshii}{2004}]{naga04}
			Nagashima M., Yoshii Y., 2004, ApJ, 610, 23

\bibitem[\protect\citeauthoryear{Pagel et al.}{1978}]{pagel:smc} Pagel
			B.~E.~J., Edmunds M.~G., Fosbury R.~A.~E.,
			Webster B.~L., 1978, MNRAS, 184, 569

\bibitem[\protect\citeauthoryear{Peletier \&
			Christodoulou}{1993}]{peletier1993} Peletier
			R.~F., Christodoulou D.~M., 1993, AJ, 105, 1378

\bibitem[\protect\citeauthoryear{Pelupessy, van der Werf \&
		Icke}{2004}]{pelu04} Pelupessy F.~I., van der Werf
		P.~P., Icke V., 2004, A\&A, 422, 55

\bibitem[\protect\citeauthoryear{Penny et al.}{2009}]{pe09} Penny S. J.,
			Conselice C. J., De Rijcke S., Held E. V., 2009,
			MNRAS, 393, 1054

\bibitem[\protect\citeauthoryear{Peterson \& Caldwell}{1993}]{peter93}
			Peterson R.~C., Caldwell N., 1993, AJ, 105, 1411

\bibitem[\protect\citeauthoryear{Revaz et al.}{2009}]{re09} Revaz Y.,
  Jablonka P., Sawala T., Hill V., Letarte B., Irwin M., Battaglia G.,
  Helmi A., Shetrone M. D., Tolstoy E., Venn K. A., 2009, A\&A, 501, 189

\bibitem[\protect\citeauthoryear{Rhode et al.}{1999}]{rhode99} Rhode
			K.~L., Salzer J.~J., Westpfahl D.~J., Radice
			L.~A., 1999, AJ, 118, 323

\bibitem[\protect\citeauthoryear{Roychowdhury et
			al.}{2010}]{roychowdhury:flatdist} Roychowdhury
			S., Chengalur J.~N., Begum A., Karachentsev
			I.~D., 2010, MNRAS, 404, L60

\bibitem[\protect\citeauthoryear{S{\'a}nchez-Janssen, M{\'e}ndez-Abreu,
			\& Aguerri}{2010}]{sanchez10}
			S{\'a}nchez-Janssen R., M{\'e}ndez-Abreu J.,
			Aguerri J.~A.~L., 2010, MNRAS, 406, L65

\bibitem[\protect\citeauthoryear{Saviane, Held, \&
			Piotto}{1996}]{saviane1996} Saviane I., Held
			E.~V., Piotto G., 1996, A\&A, 315, 40

\bibitem[\protect\citeauthoryear{Sawala et al.}{2010}]{sa10} Sawala T.,
		Scannapieco C., Maio U., White S., 2010, MNRAS, 402,
		1599

\bibitem[\protect\citeauthoryear{Sawala et al.}{2011}]{sa11} Sawala T.,
		Guo Q., Scannapieco C., Jenkins A., White S., 2011,
		MNRAS, 64

\bibitem[\protect\citeauthoryear{Schmidt}{1959}]{schmidt59} Schmidt M.,
		1959, ApJ, 129, 243

\bibitem[\protect\citeauthoryear{Skillman}{2005}]{ski05} Skillman
		E. D., 2005, NewAR, 49, 453

\bibitem[\protect\citeauthoryear{Smith Castelli et
			al.}{2008}]{smith2008} Smith Castelli A.~V.,
			Bassino L.~P., Richtler T., Cellone S.~A., Aruta
			C., Infante L., 2008, MNRAS, 386, 2311

\bibitem[\protect\citeauthoryear{Springel}{2005}]{springel05} 
		Springel V., 2005, MNRAS, 364, 1105 

\bibitem[\protect\citeauthoryear{Staveley-Smith, Davies, \&
			Kinman}{1992}]{staveley:flatdist} Staveley-Smith
			L., Davies R.~D., Kinman T.~D., 1992, MNRAS,
			258, 334

\bibitem[\protect\citeauthoryear{Stewart et al.}{2000}]{stewart00}
			Stewart S.~G., et al., 2000, ApJ, 529, 201

\bibitem[\protect\citeauthoryear{Stinson et al.}{2006}]{sti06} Stinson
		G., Seth A., Katz N., Wadsley J., Governato F., Quinn
		T., 2006, MNRAS, 373, 1074

\bibitem[\protect\citeauthoryear{Stinson et al.}{2007}]{sti07} Stinson
			G. S., Dalcanton J. J., Quinn T., Kaufmann T.,
			Wadsley J., 2007, ApJ, 667, 170

\bibitem[\protect\citeauthoryear{Sung et al.}{1998}]{sung:flatdist} Sung
			E.-C., Han C., Ryden B.~S., Patterson R.~J.,
			Chun M.-S., Kim H.-I., Lee W.-B., Kim D.-J.,
			1998, ApJ, 505, 199

\bibitem[\protect\citeauthoryear{Sutherland \& Dopita}{1993}]{suthdop}
			Sutherland R.~S., Dopita M.~A., 1993, ApJS, 88,
			253

\bibitem[\protect\citeauthoryear{Tolstoy \&
		  Saha}{1996}]{tolstoy:cmdanal} Tolstoy E., Saha A.,
		  1996, ApJ, 462, 672

\bibitem[\protect\citeauthoryear{Tolstoy et al.}{2004}]{tolstoy04}
		Tolstoy E., et al., 2004, ApJ, 617, L119

\bibitem[\protect\citeauthoryear{Tolstoy, Hill \&
			Tosi}{2009}]{tolstoy:dgreview} Tolstoy E., Hill
			V., Tosi M., 2009, ARA\&A, 47, 371

\bibitem[\protect\citeauthoryear{Tosi}{2007}]{tosi:beyondlg} Tosi M., 
2007, ASPC, 374, 221 

\bibitem[\protect\citeauthoryear{Tosi et al.}{1991}]{tosi:cmdanal} 
Tosi M., Greggio L., Marconi G., Focardi P., 1991, AJ, 102, 951 

\bibitem[\protect\citeauthoryear{Travaglio et al.}{2004}]{travaglio04}
			Travaglio C., Hillebrandt W., Reinecke M.,
			Thielemann F.-K., 2004, A\&A, 425, 1029

\bibitem[\protect\citeauthoryear{Tsujimoto et al.}{1995}]{tsujimoto95}
			Tsujimoto T., Nomoto K., Yoshii Y., Hashimoto
			M., Yanagida S., Thielemann F.-K., 1995, MNRAS,
			277, 945

\bibitem[\protect\citeauthoryear{Valcke et al.}{2008}]{sander:dgmodels}
			Valcke S., De Rijcke S., Dejonghe H., 2008,
			MNRAS, 389, 1111

\bibitem[\protect\citeauthoryear{Valcke et al.}{2010}]{sander:sph} 
		Valcke S., De Rijcke S., R{\"o}diger E., Dejonghe H.,
		2010, MNRAS, 408, 71

\bibitem[\protect\citeauthoryear{Valcke}{2010}]{sander:phd} 
  Valcke S, 2010, PhD thesis, Ghent University

\bibitem[\protect\citeauthoryear{van Zee et al.}{2001}]{vanzee01} van
  Zee L., Salzer J. J., Skillman E. D., 2001, AJ, 122, 121

\bibitem[\protect\citeauthoryear{van Zee}{2002}]{vanzee02} van Zee L.,
  2002, ASP Conference Proceedings, Vol. 285., ASPC, 285, 333

\bibitem[\protect\citeauthoryear{van Zee et al.}{2004}]{vanzee04} van
  Zee L., Skillman E. D., Haynes M. P., 2004, AJ, 128, 121

\bibitem[\protect\citeauthoryear{Vazdekis et al.}{1996}]{vazdekis96}
		Vazdekis A., Casuso E., Peletier R.~F., Beckman J.~E.,
		1996, ApJS, 106, 307

\bibitem[\protect\citeauthoryear{Walker et al.}{2007}]{walker07} Walker
		M.~G., Mateo M., Olszewski E.~W., Gnedin O.~Y., Wang X.,
		Sen B., Woodroofe M., 2007, ApJ, 667, L53

\bibitem[\protect\citeauthoryear{Walter et al.}{2008}]{walter08} 
Walter F., Brinks E., de Blok W.~J.~G., Bigiel F., Kennicutt R.~C., 
Thornley M.~D., Leroy A., 2008, AJ, 136, 2563 

\bibitem[\protect\citeauthoryear{Weaver et al.}{1977}]{weaver77} 
Weaver R., McCray R., Castor J., Shapiro P., Moore R., 1977, ApJ, 218, 377 

\bibitem[\protect\citeauthoryear{Weisz et al.}{2009}]{weisz09} 
Weisz D.~R., Skillman E.~D., Cannon J.~M., Dolphin A.~E., Kennicutt R.~C., 
Lee J., Walter F., 2009, ApJ, 704, 1538 

\bibitem[\protect\citeauthoryear{Wilkinson et al.}{2004}]{wilkinson04}
		Wilkinson M.~I., Kleyna J.~T., Evans N.~W., Gilmore
		G.~F., Irwin M.~J., Grebel E.~K., 2004, ApJ, 611, L21

\bibitem[\protect\citeauthoryear{Zucker et al.}{2007}]{zucker2007}
			Zucker D.~B., et al., 2007, ApJ, 659, L21






\end{thebibliography}
\end{document}